\documentclass[12pt,preprint]{aastex}
\usepackage{apjfonts,graphicx,graphics}
\usepackage{amsmath}
\usepackage{natbib}
\usepackage{rotating}
\newcommand{\simless}
     {\ensuremath{\lower
3pt\hbox{$\rlap{\raise5pt\hbox{$\char'074$}}\mathchar"7218$}}}
\newcommand{\simgreat}
     {\ensuremath{\lower
3pt\hbox{$\rlap{\raise5pt\hbox{$\char'076$}}\mathchar"7218$}}}
\newcommand{\simgt}{\lower.5ex\hbox{$\; \buildrel > \over \sim \;$}}
\newcommand{\simlt}{\lower.5ex\hbox{$\; \buildrel < \over \sim \;$}}
\shorttitle{Magnetic Field Significance}
\shortauthors{Koch et al.}
\begin{document}
\title{Quantifying the Significance of the Magnetic Field 
from Large-Scale Cloud 
to Collapsing Core: \\
Self-Similarity, Mass-to-Flux Ratio and Star Formation Efficiency}
\author{
Patrick M. Koch\altaffilmark{1},
Ya-Wen Tang\altaffilmark{2,3}
\&
Paul T. P. Ho\altaffilmark{1,4}
}
\altaffiltext{1}{Academia Sinica, Institute of Astronomy and
 Astrophysics, Taipei, Taiwan}
\altaffiltext{2}{Universit\'e de Bordeaux, Observatoire Aquitain des Sciences de l'Univers,
2 rue de l'Observatoire, BP 89, F-33271 Floirac Cedex, France}
\altaffiltext{3}{CNRS, UMR 5804, Laboratoire d'Astrophysique de Bordeaux,
2 rue de l'Observatoire, BP 89, F-33271 Floirac Cedex, France}
\altaffiltext{4}{Harvard-Smithsonian Center for Astrophysics, 60
 Garden Street, Cambridge, MA 02138, USA}

\email{pmkoch@asiaa.sinica.edu.tw}
%
%
\begin{abstract}
Dust polarization observational results are analyzed 
for the high-mass star formation region W51 from the 
largest parent cloud ($\sim$ 2~pc, JCMT) to the large-scale envelope 
($\sim$ 0.5~pc, BIMA) down to the collapsing core e2 ($\sim$ 60~mpc, SMA).
Magnetic field and dust emission gradient orientations reveal a correlation 
which becomes increasingly more tight with higher resolution. The previously
developed polarization - intensity gradient method \citep{koch11} is applied
in order to quantify the magnetic field significance. This technique provides
a way to estimate the local magnetic field force compared to gravity without 
the need of any mass or field strength measurements, solely making use of 
measured angles which reflect the geometrical imprint of the various forces. 
All three data sets 
clearly show regions with distinct features in the field-to-gravity force ratio.
Azimuthally averaged radial profiles of this force ratio reveal a transition 
from a field dominance
at larger distances to a gravity dominance closer to the emission peaks. 
Normalizing these profiles to a characteristic core scale
points toward self-similarity. Furthermore, the polarization
intensity-gradient method is linked to the mass-to-flux ratio, providing a new
approach to estimate the latter one without mass and field strength inputs.
A transition from a magnetically 
supercritical to a subcritical state as a function of distance from 
the emission peak is found for the e2 core.
Finally, based on the measured radius-dependent field-to-gravity 
force ratio we derive a modified star formation efficiency with a diluted gravity 
force. Compared to a standard (free-fall) efficiency, the observed field is 
capable of reducing the efficiency down to 10\% or less.

\end{abstract}
%
%
\keywords{ISM: clouds --- ISM: magnetic fields, polarization
--- ISM: individual
          (W51, W51 e2) --- Methods: polarization}

\section{Introduction}     \label{intro}

The magnetic field is being recognized as a crucial component in the 
star formation process. Evidence for its presence and significance is growing
with the advance of an increasing number of instruments capable of providing
high-quality polarization observations. To date, various observations -- based
on different emission or absorption mechanisms over a range of wavelengths --
cover scales from a few pc down to mpc in molecular clouds. Nevertheless, the precise
role of the magnetic field and its interplay with, e.g. turbulence and gravity, 
remain a debated topic in the current literature. In order to make further 
progress, knowledge of both the field morphology and the field strength needs
to be combined. It is unfortunate that many of the currently available 
observational techniques can only address one or the other of these field 
properties. Whereas dust polarization reveals a plane of sky projected 
magnetic field orientation and morphology \citep[e.g.,][]{hildebrand88}, 
it does not provide any information
about the magnetic field strength. On the other hand, Zeeman splitting in 
spectral lines \citep[e.g.,][]{crutcher09}
gives a line-of-sight field strength, but typically the field morphology
is not recovered due to the only isolated detections. An exception might
be the recent works by \citet{surcis11} and \citet{vlemmings11}, 
where the morphology starts to 
become visible based on several tens of maser detections.

Besides the intrinsic shortcomings of the observing techniques mentioned
above, analysis tools and interpretation of polarization data are still in 
a developmental stage. Consequently, the benefit of the additional polarization 
information beyond its mere detection and imaging purpose is often not very
obvious. With the growing number of high-quality polarization observations 
from various instruments, it is, thus, paramount to investigate new methods
and strategies to further explore and reveal the physics hidden in polarization
data. In order to make further progress, efforts on several fronts are needed and  
eventually must be combined. On one hand, magneto-hydrodynamic simulations
producing synthetic observational maps can characterize 
generic features and provide guidelines for data interpretation
\citep[e.g.,][]{nakamura11,falceta08,li04,allen03}. It is 
further desirable to include radiative transfer modeling in this approach. 
On the other hand, methods and techniques inspired by 
observed data can lead to new unexpected insights in a phenomenological 
way. Such recent approaches are, e.g., the polarization dispersion function 
used to trace the relative turbulence level \citep{houde09, hildebrand09} and
the ambipolar diffusion scale isolated by comparing coexisting ion and molecular 
line spectra \citep{hezareh11,li08}. Here, we are further exploring the 
recently developed
polarization - intensity gradient method \citep{koch11}. This technique
leads to a local
position-dependent measurement of the magnetic field strength, 
and it additionally provides an estimate of the local field-to-gravity 
force ratio in a model-independent way.

We are applying our method to a set of dust
emission polarization data in the sub-millimeter regime, where the
dust grains are thought to be aligned with their shorter axis
parallel to the magnetic field lines 
due to radiative torques
\citep{draine96, draine97, lazarian00}.
The emitted light, therefore, 
appears to be polarized perpendicular to the field lines
\citep[e.g.,][]{hildebrand88}.
Dust polarization observations are now being routinely carried out 
with the Submillimeter Array (SMA)
(Tang et al. 2012; in preparation;
\citet{tang10, tang09b, tang09a, rao09, girart09, girart06}).
This study is part of the program on the SMA\footnote{
The Submillimeter Array is a joint project between the Smithsonian
Astrophysical Observatory and the Academia Sinica Institute of Astronomy
and Astrophysics, and is funded by the Smithsonian Institution and the
Academia Sinica.
}
\citep{ho04} to investigate the structure of the magnetic field from large
to small scales.

The paper is organized as follows. Focusing on the magnetic field detections
in the W51 star formation region, Section \ref{source} describes the relevant 
polarization observations
from the JCMT, BIMA and the SMA. The key results
of the polarization - intensity gradient method \citep{koch11} -- which serve
as a starting point for this work here -- are summarized in Section \ref{method}.
Section \ref{results} starts with pointing out the correlation between the 
magnetic field and intensity gradient orientations, and then presents our 
results on the magnetic field significance. Implications of our findings for the 
mass-to-flux ratio and the star formation efficiency are discussed in 
Section \ref{discussion}. A summary and conclusion are given in 
Section \ref{summary}.

\section{Observations and Source Description}  \label{source}

The W51A\footnote{
The source names W51A (G49.5-0.4) or W51 are used synonymously in the 
following sections.
}
cloud is one of the most active and luminous high mass star formation
sites in the Galaxy. 
Located at a distance of $\sim$7 kpc \citep{genzel81},
1$\arcsec$ is equivalent to $\sim$ 0.03 pc.
\citet{chrys02} measured the polarization at 850$\mu$m with SCUBA on the 
JCMT across the molecular cloud at a scale of 5 pc with
a binned resolution $\theta\approx 9\farcs 3$.
Their observation encompasses two cores with polarized emission detected
both in the cores and the region in between them. 
At this scale, the polarization appears to be organized but with a morphology
that changes from the dense cores to the surrounding more diffuse areas.
\citet{lai01} reported a higher angular resolution (3$\arcsec$)
polarization map
at 1.3 mm obtained with BIMA, which resolved out large-scale structures.
In contrast to the larger scale JCMT map,
the polarization appears to be uniform across the envelope at a scale of
0.5 pc, resolving the eastern core in the JCMT map into
the sources W51 e2 and e8. 
W51 e2 is one of the strongest mm/submm continuum sources in the W51A
region.
In the highest angular resolution map
obtained
with the SMA at 870 $\mu$m with $\theta \approx 0\farcs$7,
the polarization patterns appear to be pinched in e2 and also possibly in
e8 \citep{tang09b}. In particular, the structure detected in the collapsing
core e2 reveals hourglass-like features\footnote{
Regardless of the different physical scales in the JCMT, BIMA and the SMA observations, 
areas in a any map identified by clear 
emission peaks are denoted as {\it cores}, here and in the following sections.
}
. 
These data currently provide the highest angular resolution information
on the morphology of the magnetic field in the plane of sky obtained
with the emitted polarized light in the W51A star formation site. 
A statistical analysis based on a polarization structure function (of second
order) shows a turbulent to mean magnetic field ratio which decreases
from the larger (BIMA) to the smaller (SMA) scales from $\sim 1.2$ 
to $\sim 0.7$ \citep{koch10}, possibly demonstrating that the role of magnetic 
field and turbulence evolves with scale.
Collapse signatures in W51 have been reported for 
various molecules in \citet{rudolph90, ho96, zhang97, young98, zhang98, solins04}.

Figure \ref{map_b_i} reproduces the dust continuum Stokes $I$ maps from the JCMT, 
BIMA and the SMA observations\footnote{
The reduced JCMT data are available at {\rm http://cdsarc.u-strasbg.fr/viz-bin/qcat?J/A+A/}
}, 
with their original (phase) centers at 
Right Ascension (J2000) $\alpha$=
$19^{\rm h} 23^{\rm m} 39^{\rm s}.0$,
Declination (J2000) $\delta$=$14^{\circ} 31\arcmin 08\farcs00$;
$\alpha$=
$19^{\rm h} 23^{\rm m} 44^{\rm s}.2$,
$\delta$=$14^{\circ} 30\arcmin 33\farcs4$ and
$\alpha$=
$19^{\rm h} 23^{\rm m} 43^{\rm s}.95$,
$\delta$=$14^{\circ} 30\arcmin 34\farcs00$.
Overlaid in red are the magnetic field 
segments, rotated by 90$^{\circ}$ with respect to the originally detected
polarization orientations. 
Typical measurement uncertainties of individual position angles are in the 
range of $5^{\circ}$ to $10^{\circ}$ and $3^{\circ}$ to $9^{\circ}$ for the SMA 
and for BIMA, respectively. The median uncertainty in the JCMT data is about 
$5^{\circ}$, with a few outliers between $30^{\circ}$ and $40^{\circ}$.
Additionally shown are the intensity gradient 
orientations in blue, which are relevant for the further analysis in the 
following sections.

We remark that W51A was also observed with the 350$\mu$m polarimeter 
Hertz at the Caltech Submillimeter Observatory with a resolution of 
$\sim 20 \arcsec$ \citep{dotson10}. For our purposes here, we have found these
data to be in agreement with and equivalent to the JCMT observation. 
They are, thus, not further discussed here.

\section{A Model Independent Approach: the Polarization - Intensity Gradient Method}
                                                 \label{method}

Various forces are interacting in molecular clouds. Maps of observed dust 
emission are reflecting the overall result of gravity, pressure and magnetic
field forces, and possible additional constituents. 
Dust emission and magnetic field morphologies
are left with a geometrical imprint by the combined effect of all these
forces. Interestingly, polarization orientations are often observed to be close 
to tangential to dust emission intensity contours. Therefore, magnetic field and intensity
gradient directions show a correlation, where the difference $\delta$ in their orientations
can be linked to the magnetic field strength (Figure 3 in \citet{koch11}). In this method,
various components in the 
{\bf 
ideal} magneto-hydrodynamic (MHD) force equation are 
identified in dust polarization and Stokes $I$ maps. In particular, a change
in emission intensity (gradient) is assumed to be the result of the transport
of matter driven by the combination of all the forces in the MHD equation.
Adopting this, it then follows that the gradient in the dust emission Stokes 
$I$ intensity defines the resulting direction of motion in the MHD force equation.
A force triangle, where the vector sum of all forces is set equal to the 
intensity gradient, can then be constructed (Figure 3 in \citet{koch11}).
As a result, the
total magnetic field strength as a function of position in a map
can be calculated as:
\begin{equation}
B=\sqrt{\frac{\sin\psi}{\sin\alpha}\left(\nabla P+\rho\nabla\phi\right)4\pi R},  
                                                                      \label{B}  
\end{equation}
where the angle $\psi$ is the difference in 
orientations between the gravitational
pull and the intensity gradient, and the angle $\alpha$ is the difference between
the polarization and the intensity gradient
orientations. $\rho\nabla\phi$ and 
$\nabla P$ are the gravitational pull and the pressure gradient, respectively. 
$R$ is the magnetic field radius. Generally, all variables are functions of 
positions in a map.
In the case of W51 e2, when neglecting the pressure gradient,
the field strengths vary between $\sim 2$~mG and  
$\sim 20$~mG 
with a radial profile $B(r) \sim r^{-1/2}$
\citep{koch11}.
The field strength averaged over the e2 core is $\sim$7.7~mG.

In order to determine the field strength $B$ in Equation (\ref{B}), the mass 
(gravitational potential $\phi$) and the pressure gradient, if significant, 
need to be known. Contrary to this absolute field strength measure, the relative
importance of the field compared to the other forces is directly 
imprinted in the field and intensity morphologies. With the magnetic field tension 
force $F_B=\frac{B^2}{4\pi R}$ and the 
gravitational and pressure forces $|F_G+F_P|=|\rho\nabla\phi + \nabla P|$, 
Equation (\ref{B}) is rewritten as:
\begin{equation}
\left(\frac{F_B}{|F_G+F_P|}\right)_{local}=\left(\frac{\sin\psi}{\sin\alpha}\right)_{local}
                                          \equiv \Sigma_B,
                                             \label{eq_ratio_b_g}
\end{equation}
where we have introduced $\Sigma_B$ to define the field significance.
The polarization - intensity gradient method, thus, provides a way to estimate
the local magnetic field significance relative to other forces in a model-independent way. 
It is based on a generally valid (ideal) MHD equation, but it is independent of 
any molecular cloud/core models. Furthermore, this technique to extract the field 
significance
is free of any necessity of mass and field strength measurements. The ratio in 
Equation (\ref{eq_ratio_b_g}) is solely based on measured angles which reflect
the geometrical imprint of the various forces. Consequently, this also means that
molecular clouds with accordingly scaled masses and field strengths can show 
identical morphologies. This self-similar (or scale-free) property is lifted when
calculating the field strength $B$ for a particular cloud mass in Equation (\ref{B}).

In any case, the angle factor $\frac{\sin\psi}{\sin\alpha}=\Sigma_B$
provides a quantitative criterion as to whether the magnetic field can prevent an area in a 
molecular cloud from gravitational collapse ($\Sigma_B>1)$ 
or not $(\Sigma_B<1)$.
As further demonstrated in \citet{koch11}, $\Sigma_B$
is only minimally or even not at all affected by projection effects.

\section{Results} \label{results}

\subsection{Magnetic Field -- Intensity Gradient Correlation} \label{results_b_i}

The dust continuum Stokes $I$ maps of the JCMT, BIMA and the SMA are shown in 
the left column panels in Figure \ref{map_b_i}.
In this sequence of subsequently higher resolution maps, the JCMT main core 
($\sim 40 \arcsec$ in size) is resolved into two cores ($\sim 8 \arcsec$) in the BIMA
observation, where the Northern core is further resolved into W51 e2 ($\sim 2 \arcsec$) 
in the SMA map. Overlaid on the dust continuum maps are the magnetic field segments
(red) and the intensity gradient segments (blue). For completeness, the SMA result
for W51 e2 is reproduced from \citet{koch11}. A tight correlation between 
magnetic field position angle (P.A.) and the corresponding intensity gradient P.A. 
(angle $\delta < \pi/2$ in Figure 3 in \citet{koch11}) is obvious for 
many of the pairs. The correlation coefficients -- in the definition of Pearson's 
linear correlation coefficient -- are 0.71, 0.72 and 0.95
for the JCMT, BIMA and the SMA data, respectively. Thus, there is seemingly a 
trend for an increasing alignment between magnetic field and intensity gradient
orientations with smaller scales, with the tightest correlation found for the 
collapsing core. This correlation is being investigated in a separate work 
on a larger data sample 
(Koch et al. 2012; in preparation).
It suffices to mention here that the above correlation
can possibly serve as an indicator for the role of the magnetic field through 
the evolutionary stages of a molecular cloud.

\subsection{Relative Magnetic Field Significance from Large to Small Scales}

Maps of the magnetic field to gravitational force ratio, 
$\Sigma_B=\frac{\sin \psi}{\sin \alpha}$, are displayed in the right column
panels in Figure \ref{map_b_i}. The pressure gradient force, $\nabla P$, typically 
being small compared to gravity, is omitted here.
For the 
JCMT and BIMA data, two gravity centers 
at the two emission peaks in each map are assumed in order
to calculate $\psi$. The angle $\psi$ is then simply measured in between the intensity 
gradient direction (left column panels in Figure \ref{map_b_i}) 
and the gravity center direction. For W51 e2 
(SMA data), a single gravity center at the emission peak
in the SMA map is adopted. The angle 
$\alpha=\pi/2-\delta$, where $\delta$ is the difference in between the intensity gradient 
and the magnetic field 
orientations, is deduced from the left column panels in 
Figure \ref{map_b_i}. All three maps --
originating from three different instruments, one single dish and two interferometers --
show clear differences in the ratios between core regions and areas in between the 
cores or further away. With the largest mapping area, the JCMT observation reveals very 
distinct features. With the exception of only two segments in the eastern core, both 
core regions show ratios below 0.5. In between the two cores the ratios 
reveal peaks in between 1.5 and $\sim$5,
with large stretches in the northeast-southwest direction larger than one.
The BIMA observation is rather restricted to the cores, with an area in the northern
core without polarization detection. Nevertheless, both cores clearly show ratios 
below one, except one segment in the southern core. The four segments in east-west
direction (around y-offset $\sim$ -4) reveal ratios larger or around one. Though limited
to a few segments only, this still points toward a trend of increased ratios outside
the immediate core areas. The detection by the SMA mostly resolves polarization features 
in the e2 core. The 
ratio averaged over the core is $\sim$ 0.2. The north-west extension shows two segments 
with a ratio larger than one, and an average of about 0.8. If assuming an additional
new core being formed here, the ratios get reduced to about 0.5. In any case, the 
main collapsing core very clearly shows ratios below one, whereas the more distant
north-west extension shows larger values.
We remark that the total outflow mass in e2 ($\sim 1.4~M_{\odot}$, \citet{shi10}) is 
negligibly small compared to the total core mass of about $220~M_{\odot}$ \citep{tang09b}.
Additionally, with a dust-to-gas ratio of about 1 to 100, the outflow is
unlikely to be detected in the dust continuum with the SMA sensitivity. Consequently, we 
do not expect the magnetic field and the dust continuum Stokes $I$ morphologies to be
significantly affected by outflows. Our analysis, therefore, still quantifies the field to 
gravitational force ratio.
The SMA e8 core is not further analyzed here because only a few polarization segments
are detected \citep{tang09b}.

In summary, the cores in the JCMT, BIMA and SMA data have average ratios of 0.33, 0.45
(east, west), 0.38, 0.49 (north, south) and $\sim$ 0.2 (main core), respectively. 
The other areas  reveal significantly and systematically larger ratios
($\simgt 1$). Despite probing three very different physical scales, the three data sets
equally reveal a minor role of the magnetic field, $F_B<F_G$, in the center regions, and an 
increasingly more significant field, $F_B\simgt F_G$, at larger distances.

The distinct features of the field-to-gravity force ratio in the right panels in 
Figure \ref{map_b_i} are 
further analyzed in the following. Azimuthally averaged radial profiles, 
binned at half of the beam resolution, are displayed in Figure \ref{ratio_binned}. 
The 
profiles are centered at the dust continuum peaks. 
From the JCMT data, the main (eastern) peak covering the BIMA observation and the SMA e2
core is selected. On the next smaller scale, the BIMA southern core is chosen
because its force ratio in the center has a smaller uncertainty than the northern
core.
As already manifest in the right panels in Figure \ref{map_b_i}, 
a clear difference between center and outer 
regions is seen. Both the JCMT and SMA profiles show ratios across the center
($\sim 0.3$ for JCMT and $\sim 0.2$ for SMA) with relatively little variations.
This is followed by a rather abrupt increase over a short distance (one bin)
by a factor of $\sim 2$ (JCMT) and $\sim 4$ (SMA). A plateau, with ratios $\sim 0.8$
for both JCMT and SMA, then extends over distances comparable to or larger 
than the center regions. The JCMT data, with the largest mapped area, then show 
another abrupt change to a ratio beyond one. No polarized emission is detected
for W51 e2 at larger distances. The case of BIMA is less clear, but still
shows values in the center of $\sim 0.1$ and $\sim 0.6$, followed by a plateau around
0.6 and an increase to larger than one at the largest distances. 
For comparison, azimuthally averaged emission intensity profiles are also shown.

For completeness, profiles for the JCMT western and the BIMA northern core are shown
in Figure \ref{profiles_completeness}. The result for the very central region
in the BIMA northern core is less conclusive because only a single value around 0.7 
with a $\pm 50$\% error is available for the force ratio.
Nevertheless, values are clearly below one
in the core region with a trend to grow at larger distances up to about 0.9. Similarly
to the eastern core, the western core in the JCMT data shows ratios around 0.2 to 0.6 
up to about 20$\arcsec$ in distance. Further away from the peak, ratios grow up to
about 5 with some oscillatory behavior remaining larger than one. Thus, despite being
less pronounced, the two additional cores here also reveal a transition in their 
force ratios between central and more distant ratios. In particular, in the 
overlapping region between the BIMA northern core and the SMA e2 core, the ratios
show similar trends regardless of the different field morphologies (hourglass-like
for SMA, more uniform for BIMA).

Errors in the field-to-gravity force ratio are calculated by propagating the 
measurement uncertainties $\Delta\psi$ and $\Delta\alpha$  through 
Equation (\ref{eq_ratio_b_g}). The measured magnetic field $P.A.$ uncertainty, 
in the range of a few degrees to $\sim 10^{\circ}$, is assumed for $\Delta\alpha$.
A typical uncertainty of $3^{\circ}-5^{\circ}$ in $\Delta\psi$ results after
interpolation when calculating the intensity gradients. This leads to errors for 
individual ratios in Figure \ref{ratio_binned} 
of $\sim\pm 4$\% to $\pm 19$\%, $\sim\pm 4$\% to $\pm 40$\%
and $\sim\pm 2$\% to $\pm 20$\% (with a single outlier at $\pm 60$\%) for the 
SMA, BIMA and the JCMT data, respectively. After averaging in each bin, the errors are typically
reduced due to the sample variance factor, resulting in average errors of $\sim\pm 10$\%
or less. Average errors of $\sim\pm 20$\% remain for two bins in the JCMT data.
A 10\% measurement uncertainty is conservatively estimated for the emission 
intensity when calculating
errors for its radial profiles. Binned errors are then typically at the percent level
(Figure \ref{ratio_binned}).
Except for the single central value for the force ratio in the BIMA northern core
($\sim \pm 50$\% error), similar errors are present in Figure \ref{profiles_completeness}.

Finally, it is important to keep in mind that the correlation presented in 
Section \ref{results_b_i} (based on one angle between two orientations) can be 
affected by projection effects. Generally, all values are integrated along the 
line of sight. The field significance $\Sigma_B$ presented here is much less or 
not at all affected by projection, because it is based on the ratio of two 
angles \citep{koch11}. It, nevertheless, still deals with quantities averaged
along the line of sight.

\section{Discussion}  \label{discussion}

\subsection{Self-Similar Profiles}

Given the common features in the radial profiles of the magnetic to gravitational
force ratio (Figures \ref{ratio_binned} and \ref{profiles_completeness}), 
we address here the question of self-similarity. 
We focus on the JCMT main core where the BIMA and SMA data provide higher resolution
follow-up observations. The BIMA southern core is adopted due to its better statistics.
All three data sets in Figure \ref{ratio_binned} show a rather constant ratio 
across their cores (plateau-like or
a single binned radius), followed by a plateau with a larger ratio before 
eventually the ratio grows to values larger than or around one. With the cores in the 
maps of Figure \ref{map_b_i} being clearly identified, we choose to normalize the 
distances from the peaks to units of core sizes; i.e. all profiles are aligned
to one normalized core size. The normalization 
radii, chosen to be those bins where the emission intensity profiles in Figure \ref{ratio_binned}
flatten out, are $21\farcs85$, $3\farcs42$ and $1\farcs06$ for the JCMT, the BIMA and 
the SMA data, respectively. At these radii, the emissions are $\sim 15$\% to 20\%
of the peak emissions.
Figure \ref{ratio_universal} shows the result.

Even with some uncertainty left in the normalization, it is obvious that the aligned
profiles show a close resemblance. This self-similarity suggests that the 
interplay between magnetic and gravitational force is independent of the three
different scales probed here. Thus, the analysis here quantifies the relative magnetic 
field significance from being generally dominant or comparable to gravity at distances
twice the core size to minor or negligible inside the core.
We stress that this result seems to hold generally for all the different field
morphologies analyzed here, including even the more irregular cases shown in Figure
\ref{profiles_completeness}.

\subsection{Mass-to-Flux Ratio Derived with Polarization Intensity-Gradient Method}
                                                     \label{m_f_ratio_discussion}

In this section we investigate the connection between the force ratio in 
Equation (\ref{eq_ratio_b_g}) and the mass-to-flux ratio for a molecular cloud.
The mass scale associated with the amount of magnetic flux $\Phi$ threaded by a 
self-gravitating cloud was introduced in \citet{mestel56}. The magnetic critical
mass $M_{\Phi}=\frac{\Phi}{2\pi G^{1/2}}$, with $\Phi=\pi R^2 B$ where $R$ is 
the cloud radius, defines the maximum mass that can be supported by the magnetic 
field against gravitational collapse if no other forces are present (e.g. \citet{shu99}).
This leads to the notions of magnetically supercritical clouds when $M>M_{\Phi}$, 
and magnetically subcritical clouds when $M<M_{\Phi}$. Contrary to the mass-to-flux
ratio -- which is typically applied to the entire (global) molecular cloud -- 
the force ratio in Equation (\ref{eq_ratio_b_g}) is a local
criterion comparing the field significance with gravity at a specific location
in a cloud. Neglecting pressure 
gradients and explicitly writing out the local field force and gravity force
terms,  it can be expressed as a ratio of local flux over local mass:
\begin{equation}
\frac{\sin\psi_{\ell}}{\sin\alpha_{\ell}}=\frac{\Phi^2_{\ell}}{m^2_{\ell}}\cdot \frac{m_{\ell}}{M_{\ell}} \cdot
                            \frac{1}{R_{B,\ell}} \cdot \frac{1}{R_{\ell}^4} \cdot R_{G,\ell}^2 \cdot
                            \frac{1}{4\pi^3}\frac{\eta}{G},  \label{mf_local}
\end{equation}
where the lower index $\ell$ refers to a local quantity. $M_{\ell}$ is the gravitating
mass leading to the local gravitational force $F_{G,\ell}=G\frac{m_{\ell}M_{\ell}}{R_{G,\ell}^2}$ 
acting upon a local mass element $m_{\ell}$ at a distance $R_{G,\ell}$. 
$\Phi_{\ell}$ is the flux associated
with the local mass within a flux tube of radius $R_{\ell}$ (Figure \ref{m_f_ratio_schematic}). 
$R_{B,\ell}$ is the local 
field radius. $G$ and $\eta$ are the gravitational constant and a numerical unit 
conversion factor, respectively. Equation (\ref{mf_local}) is generally valid.
However, in order to make further use of it, the local mass element
 $m_{\ell}$ and $M_{\ell}$ need to be specified. We, therefore, make the basic assumption of 
spherical symmetry. The enclosed mass within a distance $r$ from the center is then  
$M_{\ell}(r)=\int^r_0 4\pi r^{\prime 2} \rho(r^{\prime})\, dr^{\prime}\equiv N_r \bar{m_{\ell}}(r)$,
where $N_r$ is the number of local flux tubes with an average local mass $\bar{m_{\ell}}(r)$ where
both depend on the integration radius $r$. 
A {\it local mass-to-flux ratio} then results from Equation (\ref{mf_local}): 
\begin{equation}
\frac{m_{\ell}}{\phi_{\ell}}=\left(\frac{\sin\psi_{\ell}}{\sin\alpha_{\ell}}\right)^{-1/2}
                               \cdot \left(\frac{m_{\ell}(r)}{\bar{m_{\ell}}(r)}\right)^{1/2}\cdot R_{B,\ell}^{-1/2}
                               \cdot R_{\ell}^{-1}\cdot (4\pi^3)^{-1/2}\cdot \left(\frac{\eta}{G}\right)^{1/2},
                                 \label{mf_mass}
\end{equation}
where we have used $M_{\ell}(r)=N_r \bar{m_{\ell}}(r)$ with $N_r=\frac{R^2_{G,\ell}}{R^2_{\ell}}$.
We note that the above expression is still fairly general and valid for any 
density profile $\rho(r)$. In particular, since the ratio depends on the ratio of 
the mass profile over a mean mass, only the functional form of the density profile
is relevant, but not its absolute value. 
Introducing a mean local mass, $\bar{m_{\ell}}(r)$, allows us to write the local 
mass-to-flux ratio with the explicit spherical radius dependence only in the 
second term on the right hand side of Equation (\ref{mf_mass}). 
Besides the spherical symmetry assumption for the mass distribution, 
Equation (\ref{mf_mass}) is basically valid at any position in a molecular 
cloud\footnote{
We note that it is also possible to proceed directly with Equation (\ref{mf_local}).
Assuming that both $m_{\ell}$ and $M_{\ell}$ are proportional to the integrated 
dust emission with an identical conversion factor, the ratio $\frac{m_{\ell}}{M_{\ell}}$
can directly be calculated from a dust continuum emission map. In this way, it is
valid for any arbitrary cloud shape. This is identical to the approach outlined
in \citet{koch11} where a local gravity direction is derived for any cloud shape.
Equation (\ref{mf_local}), in its most general form, then leads to a map of 
local mass-to-flux ratios. However, we are here aiming at revealing changes in the 
mass-to-flux ratios with radius. Therefore, the spherical symmetry assumption and 
azimuthal averaging are appropriate (Equation (\ref{diff_mf})). In practice, we
will calculate $\frac{m_{\ell}}{M_{\ell}}$ from the dust emission maps as 
mentioned above.
}.
The average inverse of the force 
ratio decreases with larger radius (Figure \ref{ratio_universal}), and 
$\left(\frac{m_{\ell}(r)}{\bar{m_{\ell}}(r)}\right)^{1/2}\le 1$ is monotonically
decreasing for centrally peaked density profiles, with unity in the center.
The local mass-to-flux ratio in Equation (\ref{mf_mass}) is, thus, decreasing with 
larger radius from the center.  
Azimuthally averaging Equation (\ref{mf_mass})
and normalizing it to the critical mass-to-flux ratio, we get the local normalized mass-to-flux
ratio -- for individual subvolumes of fixed size $R_{\ell}$ in a cloud -- as a 
function of radius:
\begin{equation}
\left(\frac{\Delta M}{\Delta\Phi}(r)\right)_{norm}=
                    \left<\left(\frac{\sin\psi_{\ell}}{\sin\alpha_{\ell}}\right)^{-1/2}
                    \cdot R_{B,\ell}^{-1/2}\right>_r \cdot \left(\frac{m_{\ell}(r)}{\bar{m_{\ell}}(r)}\right)^{1/2}
                    \cdot \frac{R_{\ell}^{-1}}{R_0^{-3/2}}\cdot \pi^{-1/2}
                    \le \left<\left(\frac{\sin\psi_{\ell}}{\sin\alpha_{\ell}}\right)^{-1/2}\right>_r \cdot \pi^{-1/2}
                    \label{diff_mf}
\end{equation} 
where $<...>_r$ denotes azimuthal averaging at radius $r$.
When normalizing to a local flux tube, we can simply set $R_0\equiv R_{\ell}$. Furthermore, 
as found in \cite{koch11}, $R_{B,\ell}$ is roughly constant over an observed map, and 
similar to $R_{\ell}$, which we identify with the beam resolution of an observation.
With this simplification, $\frac{R_{B,\ell}^{-1/2}R_{\ell}^{-1}}{R_0^{-3/2}}\sim 1$.
Assuming the cloud gravitating mass to be proportional to the dust emission, 
$\frac{m_{\ell}}{M_{\ell}}$ is derived from the observed emission intensity 
profiles in Figure \ref{ratio_binned}.
As already indicated with the notation, Equation (\ref{diff_mf}) quantifies the 
{\it differential mass-to-flux ratio} as a function of radius.

In a next step, we derive an {\it integrated mass-to-flux ratio}, i.e. we are asking 
whether the entire cloud inside a certain radius $r$ is magnetically supercritical. 
We then have to evaluate:
\begin{equation}
\frac{M}{\Phi}(\le r)=\frac{\sum^{N_r}m_{\ell}}{\sum^{N_r}\phi_{\ell}}=M_{\ell}(r)
                       \cdot \left(\sum^{N_r}\phi_{\ell}\right)^{-1}. \label{mf_sum}
\end{equation}
$\phi_{\ell}$ can be expressed with Equation (\ref{mf_mass}). The summation is over
all local mass and flux elements within the radius $r$, where the summation limit $N_r$ depends
on $r$. After some algebra and after normalizing with the critical mass-to-flux 
ratio we find:
\begin{equation}
\left(\frac{M}{\Phi}(\le r)\right)_{norm}=\left(\sum_{i=1}^{N_b}\left<\left
        (\frac{\sin\psi_{\ell}}{\sin\alpha_{\ell}}\right)^{1/2}\cdot R_{B,\ell}^{1/2}\right>_{r_i} 
         \cdot \frac{\bar{m}^{1/2}(r_i)\cdot m^{1/2}(r_i)}{\bar{m}(r)}\cdot N_{r_i}\right)^{-1} 
          \cdot\frac{R_l^{-3}}{r^{-7/2}} \cdot \pi^{-1/2}, \label{mf_int}
\end{equation}
where the averaging $<...>_{r_i}$ is for each bin $r_i$, with $N_{r_i}$ being the number of 
subvolumes within $r_i$ and $r_{i+1}$. $N_b$ is the number of bins within $r$
(Figure \ref{m_f_ratio_schematic}).
Equation (\ref{mf_int}) states that the integrated mass-to-flux ratio is calculated
by adding bin-averaged force ratios which are weighted with a mass function and $N_{r_i}$, 
where the latter one results from the integrated volume growing with $r$.
The normalization is with respect to $r$, i.e. the increasingly larger cloud 
(with growing radius $r$) is normalized with its corresponding critical mass-to-flux ratio. 
$R_{\ell}$ is constant and again set by the beam resolution. $R_{B,\ell}$ is also roughly 
constant. When omitting the summation, setting $N_{r_i}\equiv 1$ and $r\equiv R_{\ell}$
and replacing $\bar{m}(r)$ with $\bar{m}(r_i)$, Equation (\ref{mf_int}) reduces to the 
differential mass-to-flux ratio in Equation (\ref{diff_mf}).

Finally, a {\it global mass-to-flux ratio} for the entire cloud as a single entity can
be derived from Equation (\ref{mf_mass}) in the limiting case where all (small) 
local quantities grow to cloud-size: $m_{\ell}\rightarrow M$ and 
$<R_{B,\ell}>\approx <R_{\ell}>\approx <R_G> \equiv R$. The force ratio needs
to be averaged over the entire cloud: $\frac{\sin\psi_{\ell}}{\sin\alpha_{\ell}}\rightarrow 
\left<\frac{\sin\psi_{\ell}}{\sin\alpha_{\ell}}\right>$.
This then leads to the global mass-to-flux ratio, normalized to the critical mass-to-flux
ratio with $R_0$:
\begin{equation}
\left(\frac{M}{\Phi}\right)_{norm}=\left<\left(\frac{\sin\psi_{\ell}}{\sin\alpha_{\ell}}\right)^{-1/2}\right>
                                  \cdot \left(\frac{R}{R_0}\right)^{-3/2}\cdot \pi^{-1/2},
                                  \label{mf_glob}
\end{equation}
where $R_0\equiv R$ is the cloud radius.

The set of Equations (\ref{diff_mf}), (\ref{mf_int}) and (\ref{mf_glob}) defines a 
differential, integrated and global mass-to-flux ratio, respectively, which can 
readily be calculated from observations. These equations provide the criteria 
to evaluate whether a subvolume (local or integrated up to a certain radius) 
or an entire cloud 
is magnetically supercritical. In all cases, the inverse square root of the force
ratio is the key measure leading to an estimate independent of any mass or field 
strength input. Except for the global mass-to-flux ratio, the local ratios need the functional
form of the mass profile as an additional input. 

A schematic illustration of the different subvolumes 
analyzed by the various ratios is shown in Figure \ref{m_f_ratio_schematic}.
Figure \ref{m_f_ratio} illustrates these mass-to-flux ratios for the SMA observation.
The differential mass-to-flux ratio shows a clear transition from a magnetically
subcritical state at larger radii to a supercritical state at smaller radii.
This might indicate that a fixed subvolume can only collapse closer to the center, but not
in the outer parts of a cloud/core. This finding possibly also provides an 
explanation for why fragmentation and clustering of stars preferentially occur
at the center but not at larger radii. 
Such a result is further consistent with the theoretical work by \citet{shu97}, 
where the dilution of self-gravitational forces by the magnetic tension force
(see Section \ref{discussion_sf_eff} and Figure \ref{ratio_universal}) is found
to reduce the tendency of a disk to fragment. As further utilized in 
Section \ref{discussion_sf_eff}, this dilution is more significant at larger
radii. Consequently, the fragmentation process is more significantly suppressed here.
The number of individual collapsing subvolumes in the center region (with smaller or 
non-existent dilution) might then be linked to the multiplicity of stars. 
Additionally, the finding here might give support to the suggestion by \citet{mouschovias76}
that magnetic tension might be able to hold back the envelope and, therefore, 
prevent it from following the cloud core collapse. In a related later work, \citet{shu04}
investigated whether magnetic tension can define the masses of forming stars
by holding up the subcritical envelope of a molecular cloud while its supercritical
core is collapsing. A split monopole -- as it is 
possibly seen in the field morphology in the e2 core \citep{tang09b} -- leads to 
magnetic levitation in their work. However, they find that magnetic suspension alone 
can not keep the subcritical envelope from falling onto the center, but
additional dynamic levitation by winds and jets is needed for a halt of infall.
The results here are also in qualitative agreement with the recent numerical simulation
of giant molecular clouds with ambipolar diffusion by \citet{vazquez11}. 
They find that the mass-to-flux
ratio inside a cloud exhibits large spatial fluctuations, spanning an order of 
magnitude or more. Thus, local clumps can become magnetically supercritical 
within a globally subcritical system. Estimating $\mathcal{R}$ (the ratio between
the mass-to-flux ratio in the core and in the envelope as introduced in 
\citet{crutcher09}) from the values at the largest distances and the values
close to the center (Figure \ref{m_f_ratio}), we find $\mathcal{R}>1$. 
This would be in favor of ambipolar diffusion driven models. However, the
recent statistical analysis of $\mathcal{R}$ based on numerical simulations
with supersonic magnetohydrodynamical turbulence \citep{bertram11} yields 
a large scatter in $\mathcal{R}$ with values both smaller and larger 
then unity. It, thus, has to be acknowledged that the interpretation of 
$\mathcal{R}$ as a measure to distinguish between ambipolar diffusion and 
turbulence driven star formation is possibly not unambiguous. 

The integrated mass-to-flux ratio in Figure \ref{m_f_ratio} is driven by the 
increasing volume. It, therefore, 
grows with radius. As expected, this ratio is generally larger than the differential
ratio of a subvolume, because mass scales with volume whereas flux scales with area. 
The slight mismatch in the innermost bin results from different weightings
for the two ratios. Finally, the global mass-to-flux ratio is close to the
integrated ratio at large radii. The difference here results from the incomplete
sampling of $N_{r_i}$ in azimuth for those bins which contain some depolarization zones.
It is important to remark that both the integrated and 
global ratios present average values for the entire cloud. Therefore, they are 
likely biased toward some subvolumes. Whereas these ratios provide a description
as to whether a cloud as an entity is in a sub- or supercritical state, they 
can not properly assess the state of individual subvolumes. This question can 
only be further addressed with the differential mass-to-flux ratio. The finding
here clearly demonstrates the differences and the need for a measurement of 
the local field significance.

\subsection{Dynamical Implication and Star Formation Efficiency} \label{discussion_sf_eff}

In this section we aim at investigating further implications of the local force 
ratio $\frac{\sin\psi}{\sin\alpha}$ (Figure \ref{ratio_universal}) and the 
differential mass-to-flux ratio (Figure \ref{m_f_ratio}). We limit the discussion
here to the collapsing core W51 e2. Global and/or average cloud properties are 
typically used to characterize its state; e.g. in order to determine whether a 
system is gravitationally bound, or whether it can collapse or not. With the 
{\it local criteria} -- the magnetic field to gravity force ratio in 
Equation (\ref{eq_ratio_b_g}) and the differential mass-to-flux ratio in 
Equation (\ref{mf_mass}) -- we have tools to go one step further. We can not
only address the question whether a cloud collapses or not, but we can even
ask: {\it Where} does a cloud collapse? {\it How} does it collapse?
Does {\it all} the gas take part in the collapse? And finally, depending on 
the answers to these questions, what are the consequences for the inferred star
formation efficiency?

The change with radius in the magnetic field to gravity force ratio (Figure \ref{ratio_universal})
clearly demonstrates that gravity is not everywhere equally efficient to 
initiate and to keep driving a collapse. The result in Figure \ref{ratio_universal}
suggests and quantifies an effective gravitational force which is reduced by 
the presence of the magnetic field. Indeed, on the theory side, a concept of 
{\it diluted gravity} (by
the magnetic field) was put forward by \citet{shu97}. Thus, we can define an 
effectively acting gravitational force $F_{\ast}$ as
\begin{equation}
F_{\ast}(r)=\left(1-\left<\frac{\sin\psi}{\sin\alpha}\right>_r\right)\cdot F_{max}(r)
           \equiv<\epsilon_G>_r \cdot F_{max}(r), \label{eff_g}
\end{equation}
where $<...>_r$ denotes azimuthal averaging at radius $r$ and $F_{max}(r)$ is the maximum
non-diluted gravitational force resulting from an enclosed spherically symmetric mass
distribution  within $r$. 
$F_{\ast}$ is only defined where $\frac{\sin\psi(r)}{\sin\alpha(r)}\le 1$, which 
possibly holds only for a limited area. In Equation (\ref{eff_g}) we have 
introduced the gravity efficiency factor $\epsilon_G \in [0,1]$\footnote{
Unlike the gravity efficiency factor $\epsilon_G$, the field significance $\Sigma_B$
in Equation (\ref{eq_ratio_b_g}) is not constrained to an upper limit of one.
}
.

Besides the diluted gravity which slows down the accretion and collapse process, 
the fraction of volume of a cloud (or core) that actually can collapse is relevant;
i.e. where is the cloud/core magnetically supercritical? The differential mass-to-flux
ratio (Figure \ref{m_f_ratio}) provides a local criterion that answers this question.
The increasing ratio toward the center indicates that only the central part
of the core has accumulated enough mass to overcome the magnetic flux. By reading 
off the radius where the core state changes from magnetically sub- to supercritical, 
the potentially collapsing volume and its associated mass can be estimated. This 
defines a volume efficiency, $\epsilon_V \in [0,1]$, as compared to the maximum 
efficiency  $\epsilon_V \equiv 1$ where the entire core can collapse.

Based on the two above described findings, we proceed to estimate a star formation 
efficiency with the assumptions: (1) the available effective gravitational force
is diluted by the presence of the magnetic field and (2) only the regions where
the differential mass-to-flux ratio is larger than one will eventually collapse 
and form stars. We choose to reference our estimate to the well-known pressure-less
free-fall collapse of a spherical gas sphere. The dynamics in this case are governed
by the momentum equation $\frac{Dv}{Dt}=-G\frac{M_r}{r^2}$, where the convective 
derivative is $\frac{Dv}{Dt}=\frac{\partial v}{\partial t}+v\frac{\partial v}{\partial r}$
for the gas infall velocity $v$. The enclosed mass within radius $r$ is 
$M_r=\int_o^r 4\pi r^{\prime 2} \rho(r^{\prime})\, dr^{\prime}$ with the gas 
density $\rho$. With the diluted gravity, the momentum equation reads:
\begin{equation}
\frac{Dv}{Dt}=-<\epsilon_G>_r \cdot \frac{G M_r}{r^2}.   \label{eq_dyn}
\end{equation}
Technically, the gravity efficiency factor $<\epsilon_G>_r$ could be absorbed into an effective
density $\rho_{\ast}(r)$. Therefore, the general solution (for any $M_r$) 
of the Equation (\ref{eq_dyn})
for a fluid element starting at a distance $r_0$ and reaching the center at $r=0$ at 
the time $t$ is still valid. The time $t_{\ast}$ in the presence of the magnetic 
field is then\footnote{
For a uniform density and $<\epsilon_G>_r\equiv 1$, Equation (\ref{eq_t}) describes 
a synchronized collapse with the well-known free-fall time $t_{ff}=\sqrt{\frac{3\pi}{32 G \rho}}$.
More realistic density profiles with a central concentration, e.g. 
$\rho(r)=\rho_c\left(\frac{r}{r_c}\right)^{-\alpha}$ lead to a collapse time
$t=\sqrt{\frac{(3-\alpha)\pi}{32 G \rho_c}}\cdot \left(\frac{r_0}{r_c}\right)^{\alpha/2}$
which increases with larger distance $r_0$ from the center
and then reproduces an inside-out collapse.
}
:
\begin{equation}
t_{\ast}=\frac{\pi}{2}\sqrt{\frac{r_0^3}{2 G M_r(\rho(r),<\epsilon_G>_r)}}
                                 \approx<\epsilon_G>^{-1/2}\cdot t   \label{eq_t}
\end{equation}
Strictly speaking, $<\epsilon_G>_r$ in the above equation is still a function
of radius. In writing the right hand side we have assumed it to be averaged within
an appropriate radius.
The average accretion rate $<\dot{M}>$, assuming the entire mass $M$ within a 
cloud/core to be accreted and to collapse, is $<\dot{M}>=M/t$. Adding the volume 
efficiency factor $\epsilon_V$ together with Equation (\ref{eq_t}), the effective 
accretion rate $<\dot{M}_{\ast}>$, corrected for our observed magnetic field features,
becomes:
\begin{equation}
<\dot{M}_{\ast}>=\frac{M_{\ast}}{t_{\ast}}
\approx\frac{\epsilon_V\cdot M}{<\epsilon_G>^{-1/2}\cdot t}.    \label{eq_acc}
\end{equation}
The resulting star formation efficiency relative to the standard free-fall one is then 
expressed as:
\begin{equation}
\frac{<\dot{M}_{\ast}>}{<\dot{M}>}\approx\epsilon_V \cdot <\epsilon_G>^{1/2}. \label{eq_sf_eff}
\end{equation}
We note that the above phenomenological modeling is still valid in the presence of 
additional force terms in the momentum equation with any density profiles. 
The pressure-less free-fall collapse
simply allows us to analytically express the results.
Figure \ref{sf_efficiency} shows the reduced star formation efficiency taking into account
the effects of the 
magnetic field. The red dots indicate efficiencies estimated from the gravity dilution 
and the reduced collapse volume based on the Figures \ref{ratio_universal} and \ref{m_f_ratio}.
Since $1-\frac{\sin\psi(r)}{\sin\alpha(r)}\sim 0$ at large distances, an arbitrarily low star formation 
efficiency would result from adopting these values for $\epsilon_G$. Instead, we adopt
values averaged over the entire core (force ratio $\sim 0.33)$, the central core only ($\sim 0.2$) 
and the outer plateau ($\sim 0.85$). We consider these values adequate to estimate and sample
the gravity dilution for W51 e2. Similarly, for the volume efficiency factor we read 
a radius of about $1\arcsec$ within which the core is magnetically supercritical as compared
to an overall size of about $2\arcsec$. For comparison, a larger 
volume of about $1.5\arcsec$ is considered as well. The resulting efficiencies are reduced
at least to a conservative value of $\sim 0.35$ or less when still assuming the larger 
volume to collapse. The efficiencies drop to $\sim 0.1$ with the smaller volume, 
with a likely limit of about $\sim 0.05$ in combination with the largest observed gravity dilution.
These values are close to the observationally inferred efficiencies of a few percent
\citep[e.g.,][]{krumholz07}.
For comparison, recent numerical simulations by \citet{nakamura11} find a star formation
efficiency per global free-fall time of $\sim$ 20 to $\sim$ 30\% depending on the magnetic 
field strength. When they further take into account the feedback from outflows, their efficiencies
can be reduced by another order of magnitude.

We finally remark that the discussion here is limited to the core e2. A more complete
picture, estimating the star formation efficiency starting from the largest scales
of the cloud envelop, will need to dynamically link the observed self-similar profiles
from Figure \ref{ratio_universal}. 
Caution is needed here because linking structures of different size scales is non-trivial
due to additional structures possibly generated by, e.g. fragmentation and turbulence.
Nevertheless, this first estimate further solidifies
the significance and impact of the magnetic field. Moreover, it demonstrates that the 
magnetic field is capable of reducing the star formation efficiency by at least one 
order of magnitude or even more.

\section{Summary and Conclusion}  \label{summary}

We have applied the polarization - intensity gradient method \citep{koch11} to a 
set of single dish and interferometer dust continuum data of the W51 star formation region.
These observations cover scales from the initial parent
cloud ($\sim$ 2~pc, JCMT) to the large-scale envelope ($\sim$ 0.3~pc, BIMA)
and to the collapsing core ($\sim$ 60~mpc, SMA). Besides leading to a magnetic 
field strength as a function of position in a map, our new method also provides a 
way to estimate the local magnetic field to gravity force ratio. The technique is
model-independent, without any input from mass or field strength, solely making use
of measured angles in dust polarization and Stokes $I$ map. Here, we have focused on this
force ratio and on some of its implications. Our main results are summarized in the 
following.

\begin{figure}[t!]
\begin{center} 
\includegraphics[scale=0.48]{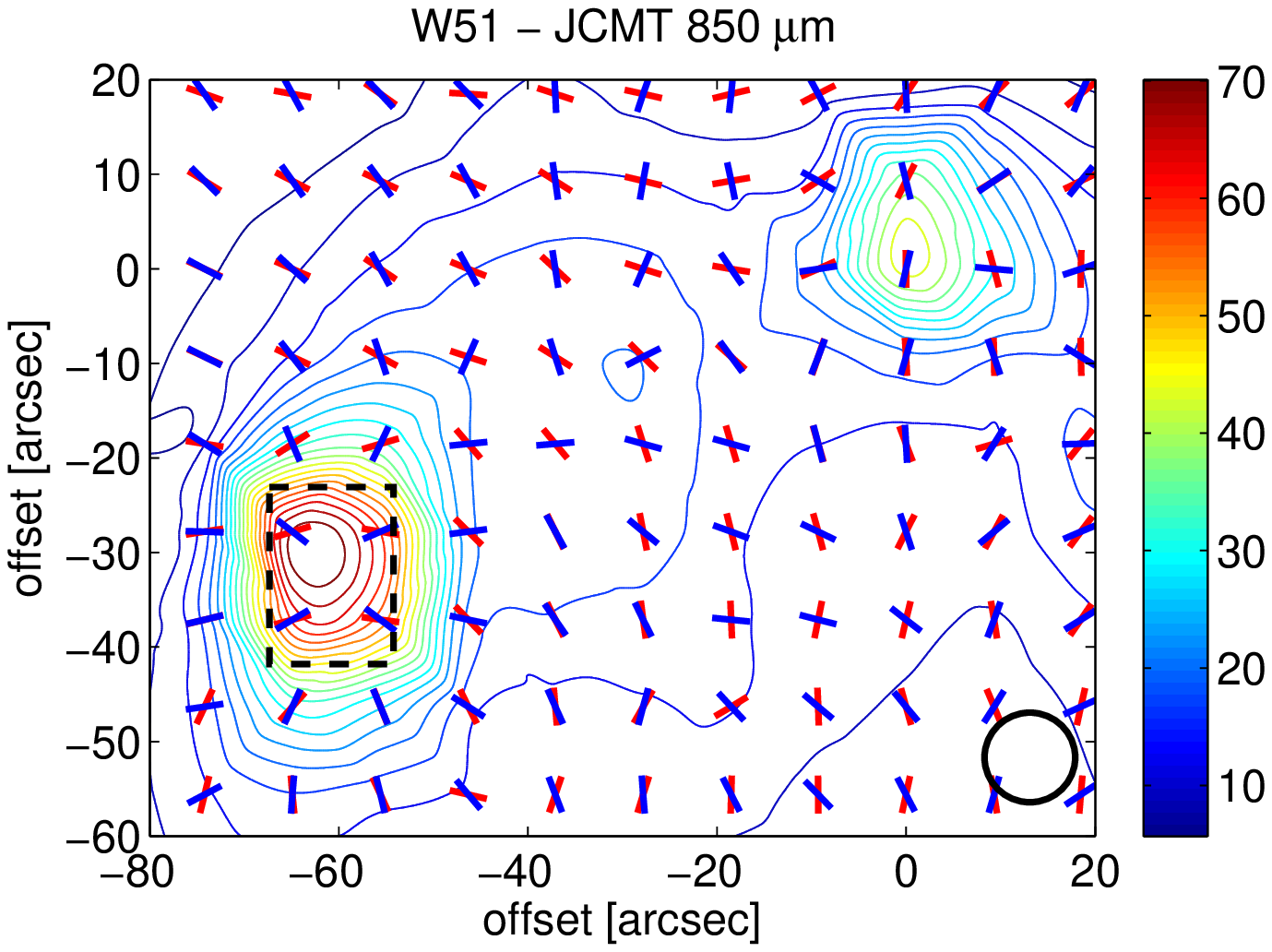}
\includegraphics[scale=0.48]{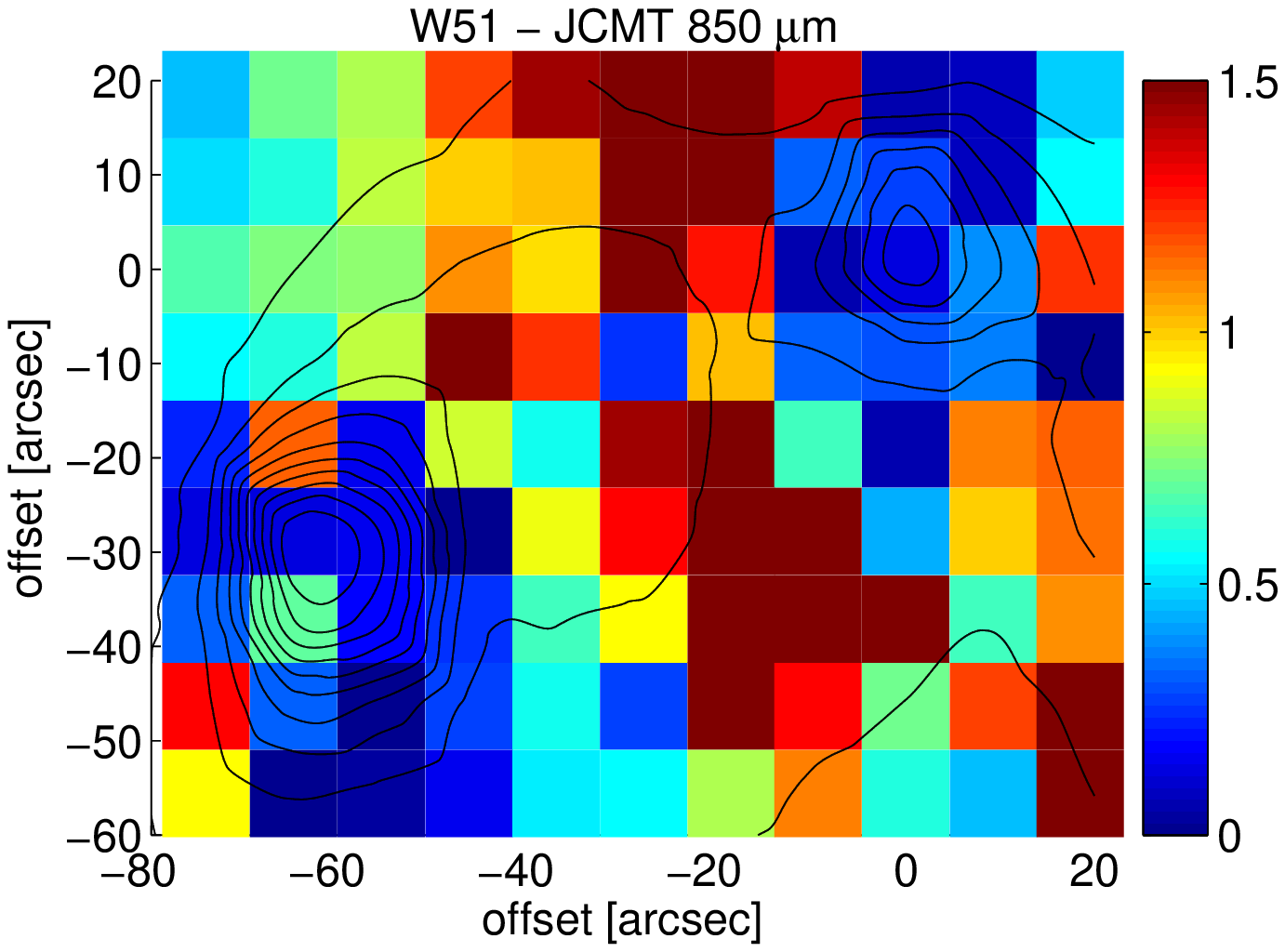}
\includegraphics[scale=0.48]{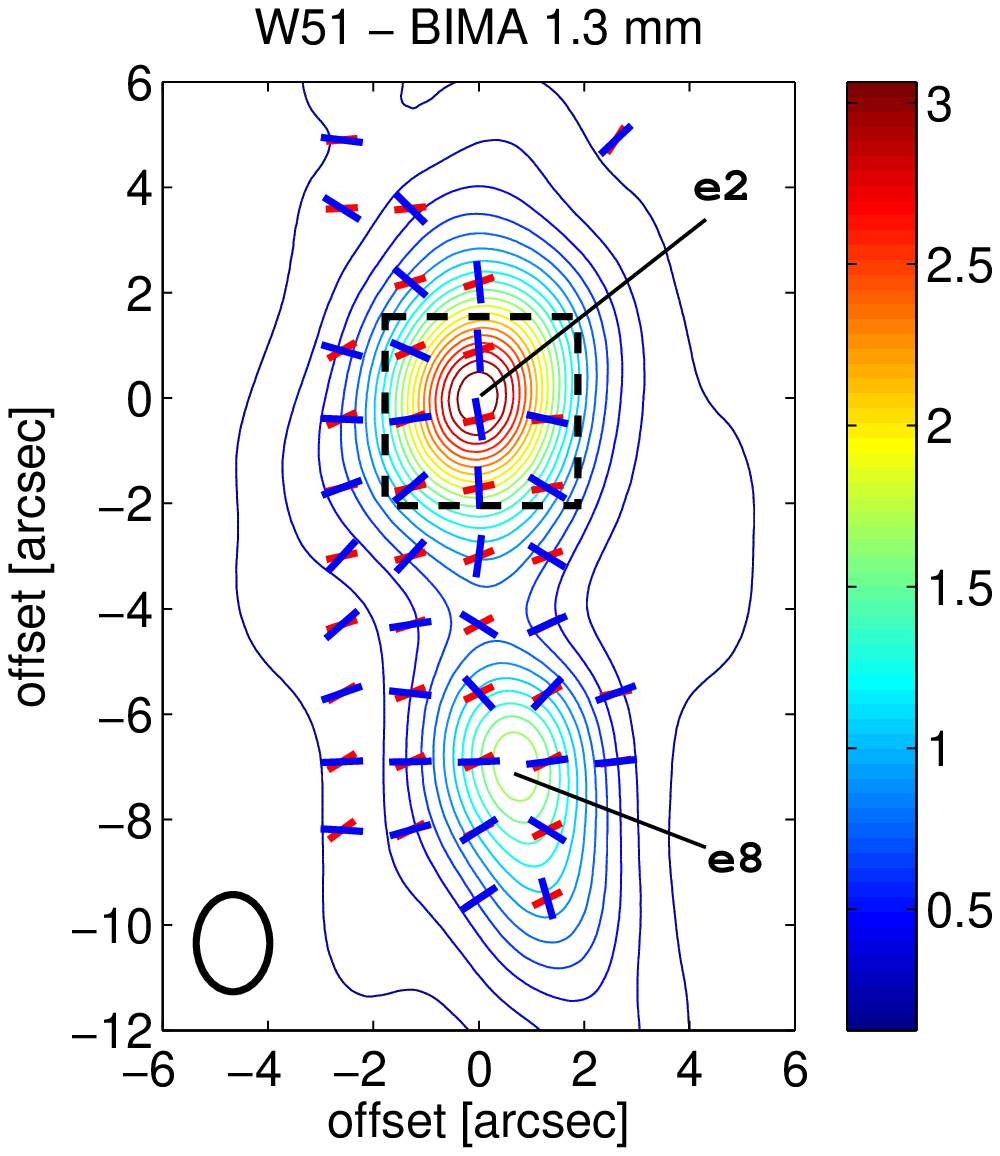}
\includegraphics[scale=0.48]{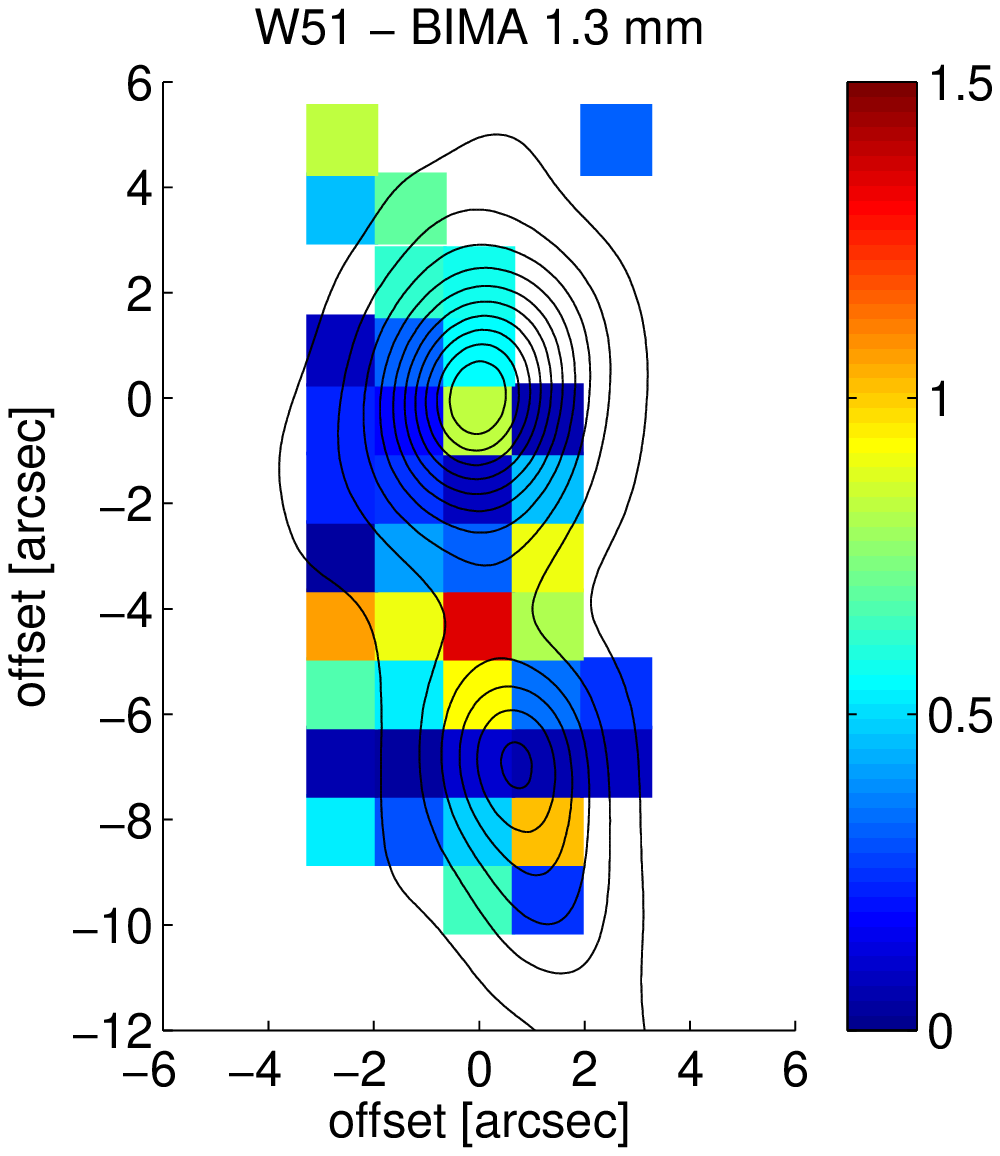}
\includegraphics[scale=0.48]{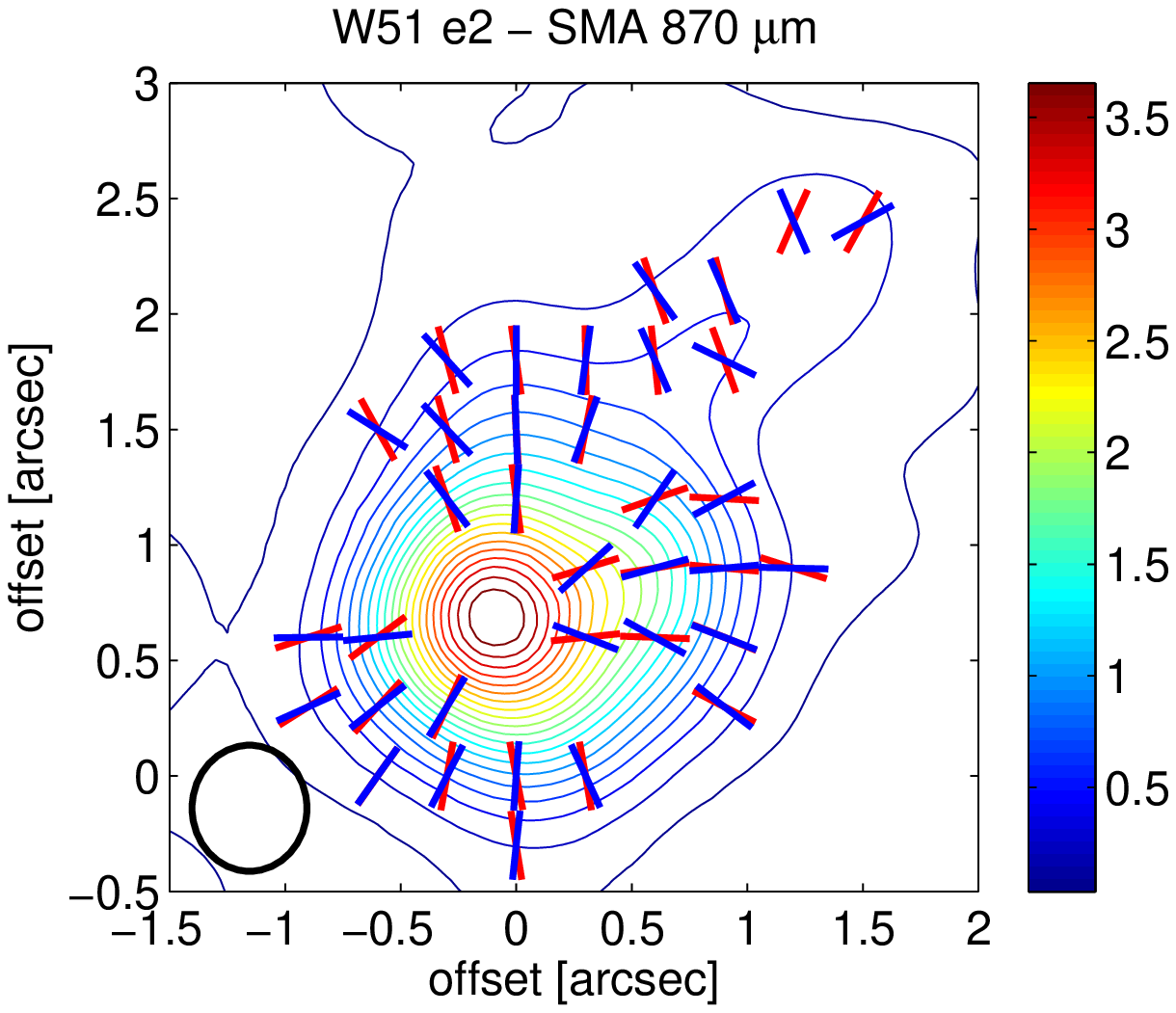}
\includegraphics[scale=0.48]{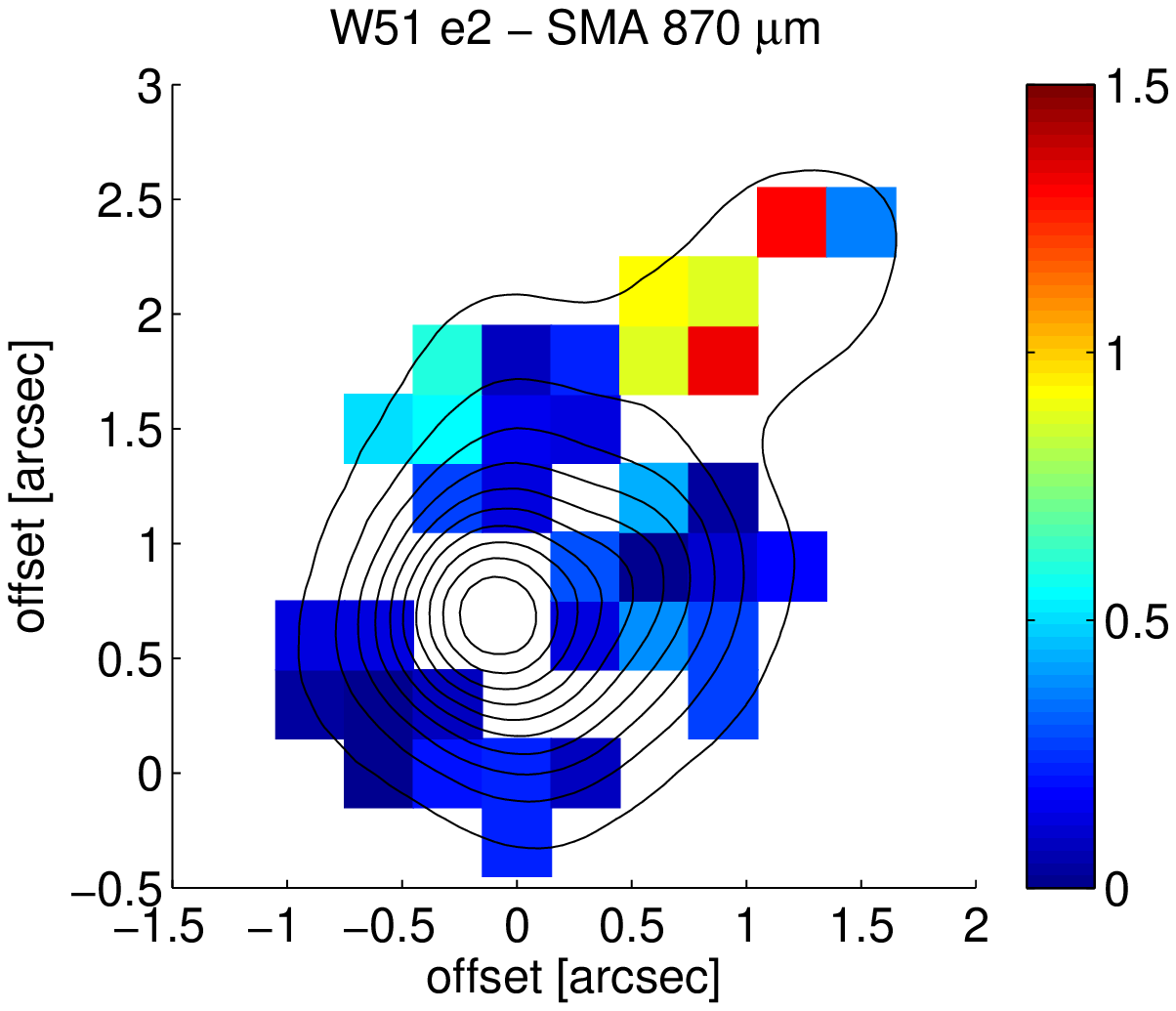}
 \caption{\label{map_b_i}\scriptsize 
Left column panels: Dust continuum Stokes $I$ observations toward W51.
From top to bottom: JCMT at 850 $\mu m$ ($\theta\approx 9\farcs3$,
from \citet{chrys02}), BIMA at 1.3~mm ($\theta\approx 3\farcs0$,
from \citet{lai01}) and SMA at  870 $\mu m$
($\theta\approx 0\farcs7$, from \citet{tang09b}).
The beam resolutions are shown with the black ellipses in the lower
left or right corners. The BIMA observation is zooming in onto the main
core (dashed rectangle) in the JCMT map. The BIMA main core (dashed square) 
is further 
resolved in the SMA map.
The cores e2 and e8 are labeled.
Colors correspond to the color
wedges on the right hand side with units in Jy/beam.
Overlaid are the magnetic field segments (red segments) 
and the intensity gradient segments (blue segments)
at the locations where polarized 
emission was detected. Magnetic field segments are plotted by rotating the 
polarization segments by 90$^{\circ}$.
The difference in $P.A.$s between the magnetic field and the intensity
gradient is the angle $\delta$ (Section \ref{results_b_i}).
The length of the segments is arbitrary, and for visual guidance only.
The axes offset positions (in arcsec) are with respect to the original 
(phase) centers of the observations.
At the distance of W51, 1$\arcsec$ is equivalent to about 30~mpc.
Right column panels: The ratio of the magnetic field force $F_B$ 
compared to the gravitational force $F_G$, 
$\Sigma_B=\frac{\sin\psi}{\sin\alpha}$,
as a function of position. 
The pixel scale is identical to the gridding in the left column panels.
From top to bottom are shown the maps for the 
JCMT, BIMA and the SMA observations. Overlaid in black are the 
Stokes $I$ dust continuum emission contours from Figure \ref{map_b_i}. 
Colors correspond to the color
wedges on the right hand side, displaying the dimensionless ratio. 
Blue colors indicate where gravity dominates over the magnetic field
($F_B<F_G$).
}
\end{center}
\end{figure}

\begin{figure} 
\begin{center} 
\includegraphics[scale=0.55]{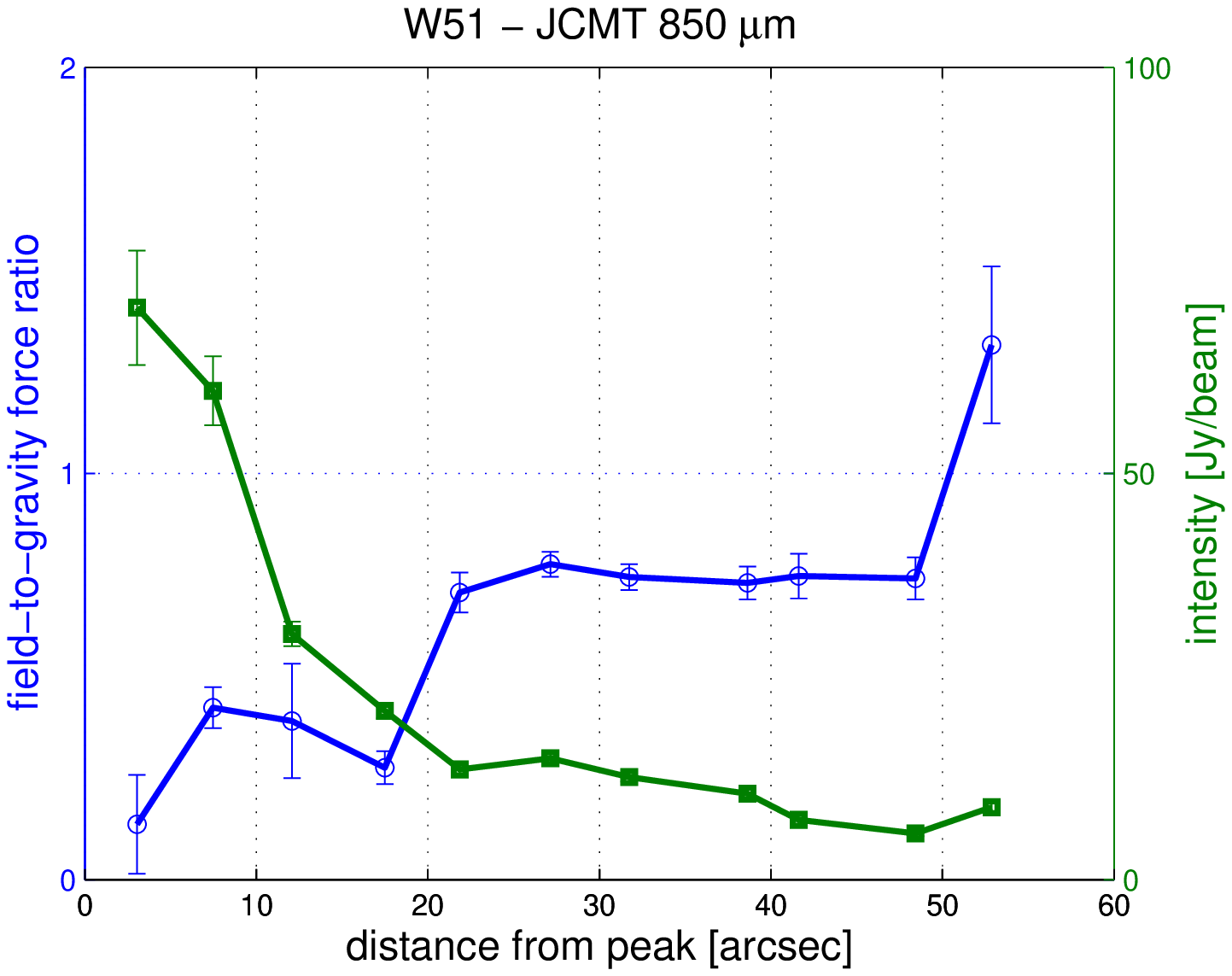}
\includegraphics[scale=0.55]{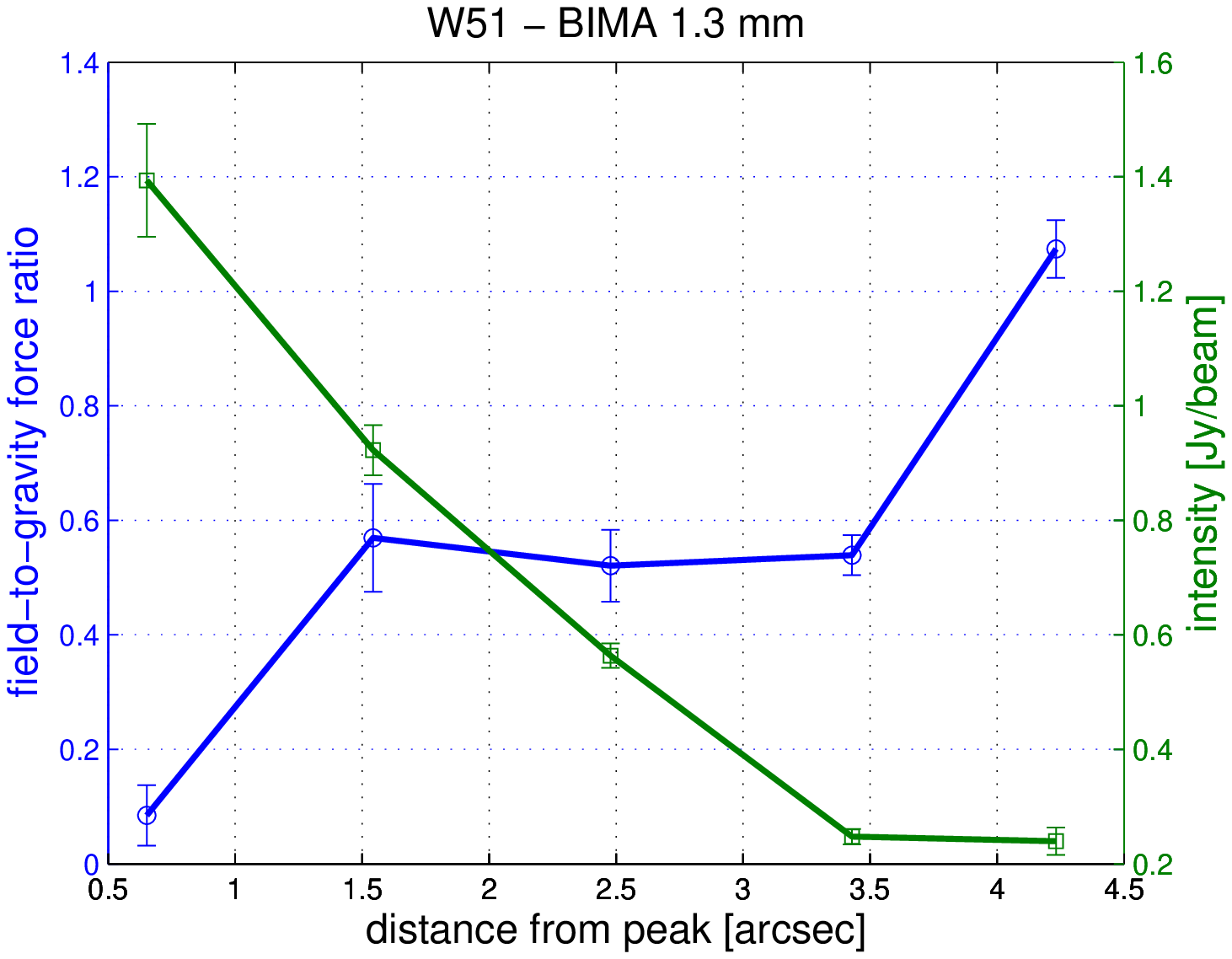}
\includegraphics[scale=0.55]{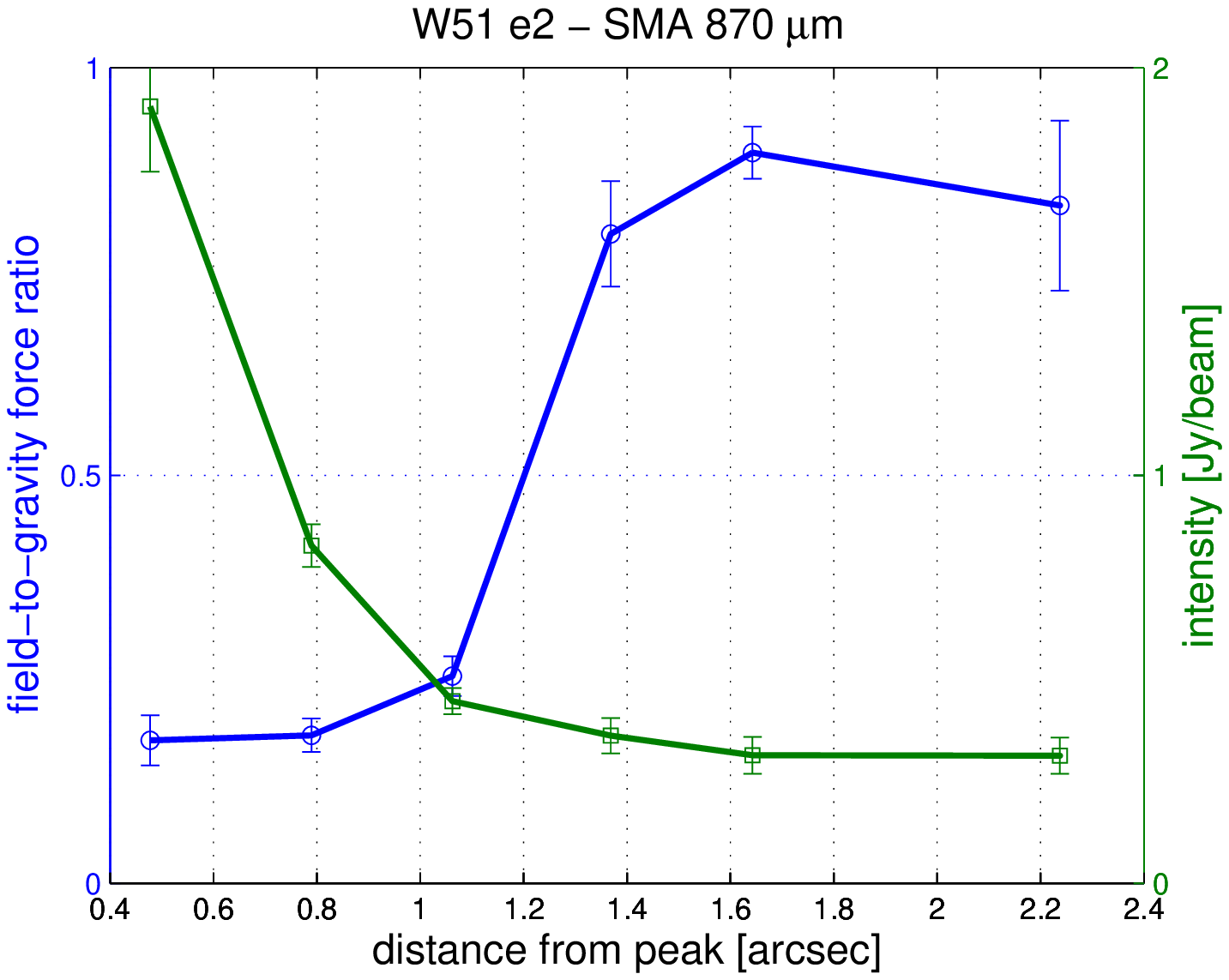}
 \caption{\label{ratio_binned}\scriptsize
The magnetic field to gravitational force ratio, $\Sigma_B=\frac{\sin\psi}{\sin\alpha}$,
as a function of distance from the emission peaks (blue). 
From top to bottom shown are the JCMT, BIMA and the SMA results.
The corresponding intensity emission profiles are displayed in green
with the axis at the right hand side.
Values are azimuthally averaged and binned at half of the beam resolution. 
From the JCMT and BIMA observations, the results for the main peak and the southern
peak, respectively, are shown.
Binned values are connected with lines for visual guidance only.
Errors for the force ratios are calculated by propagating typical measurement
uncertainties in $\psi$ and $\alpha$ through Equation (\ref{eq_ratio_b_g}).
A $10$\% uncertainty in the intensity emission is assumed. 
Error bars at the smallest and largest scales are typically growing because
of a smaller sample variance factor. 
Binned errors are all around $\sim \pm 10$\% or less, with the only 
exception being the JCMT data at the largest scale with an error of 
$\sim \pm 20$\%.
}
\end{center}
\end{figure}

\begin{figure} 
\begin{center} 
\includegraphics[scale=0.55]{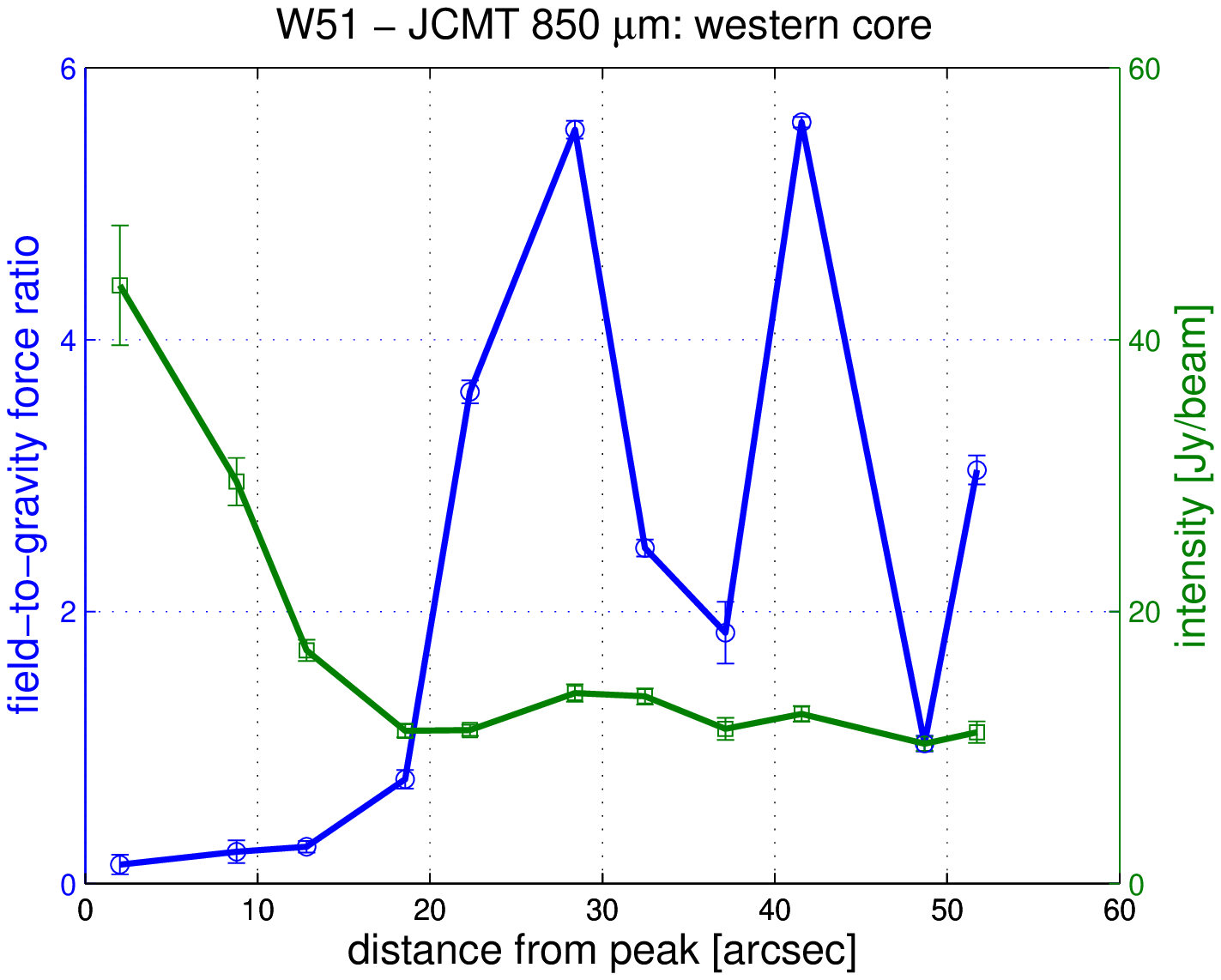}
\includegraphics[scale=0.55]{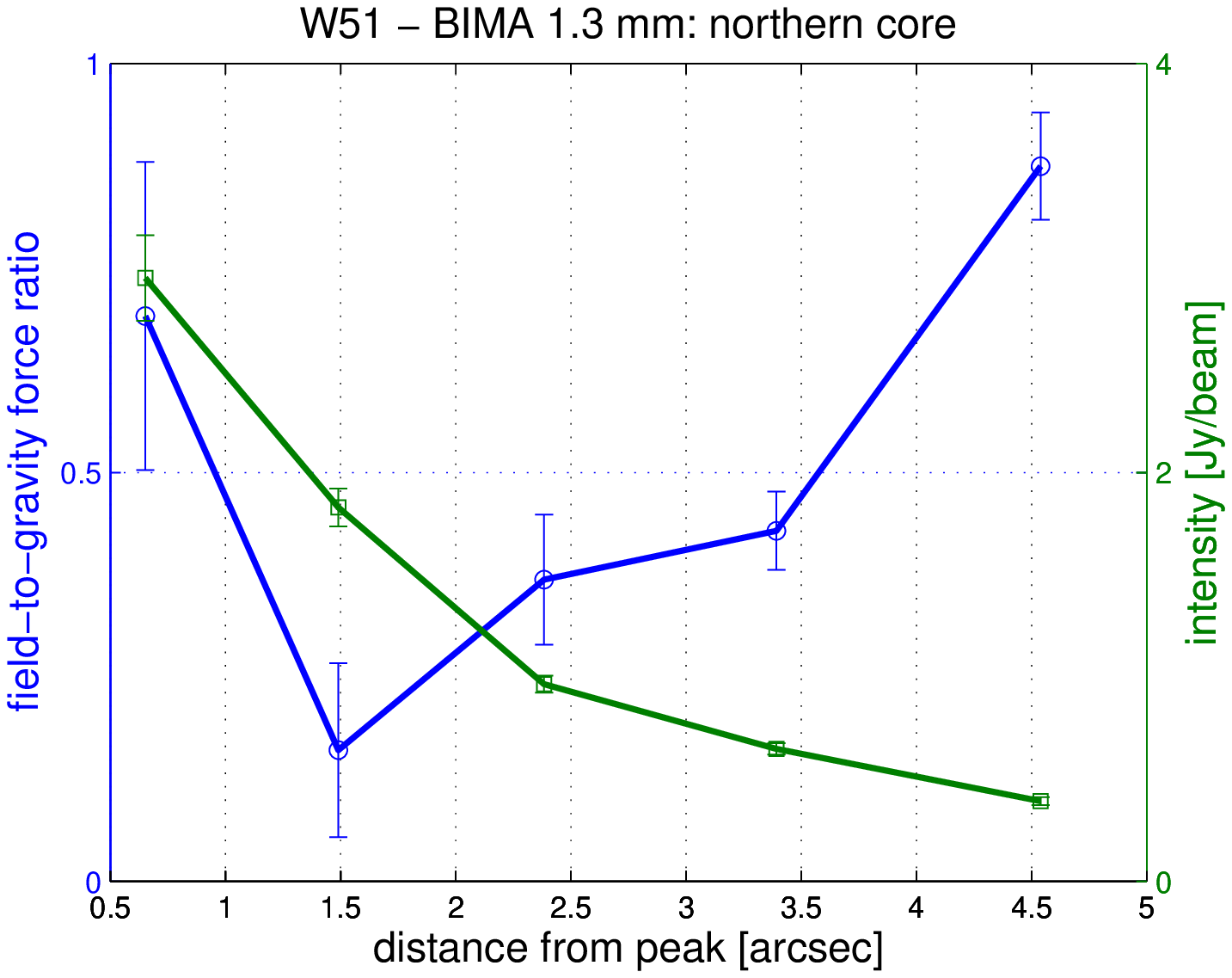}
 \caption{\label{profiles_completeness}
As in Figure \ref{ratio_binned}, the magnetic field to gravitational 
force ratio with the corresponding
intensity emission profile as a function of distance from the emission 
peak for the JCMT western and the BIMA northern core.
}
\end{center}
\end{figure}

\begin{figure} 
\begin{center} 
\includegraphics[scale=0.8]{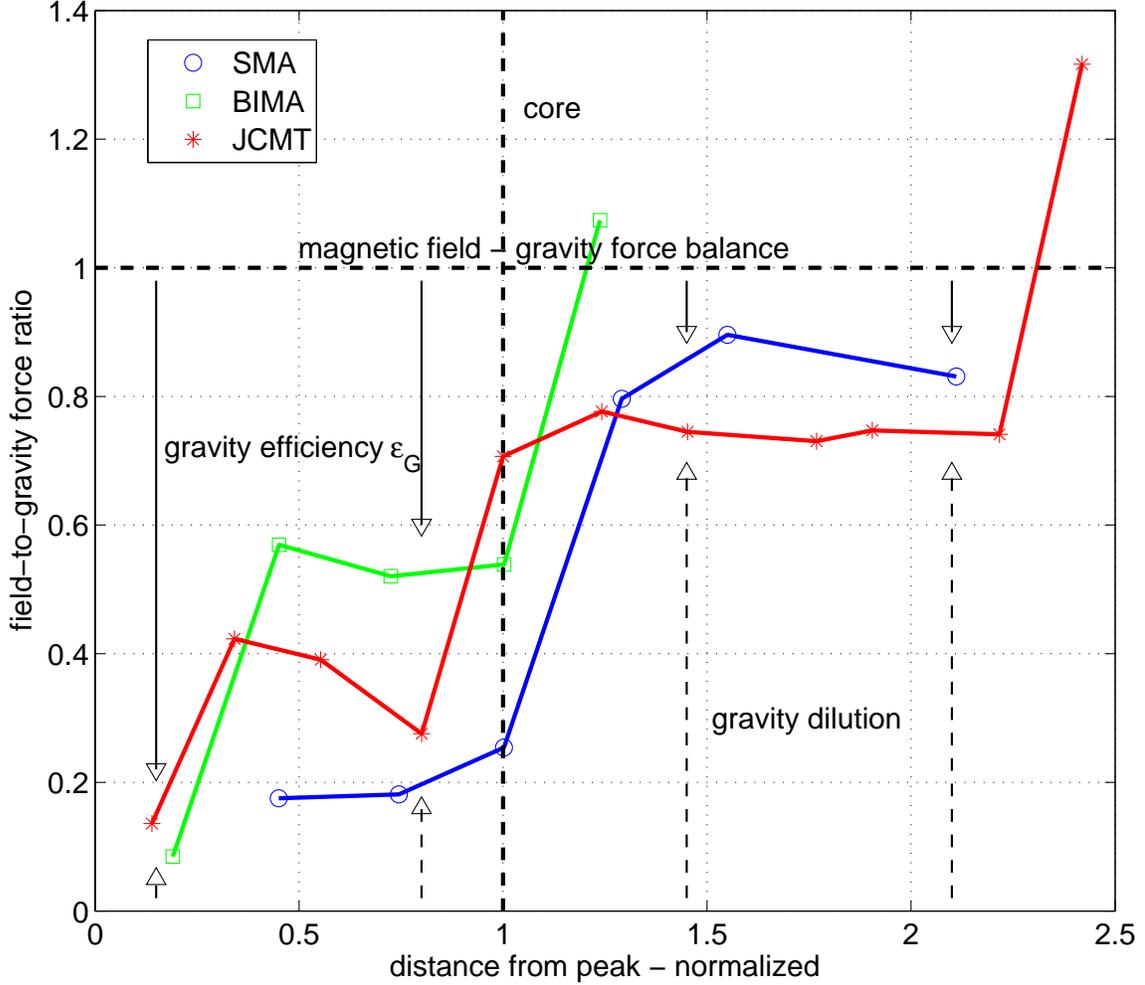}
 \caption{\label{ratio_universal}
The magnetic field to gravitational force ratio, as in Figure \ref{ratio_binned},
but with the distances from the peaks normalized to units of the core sizes.
With respect to Figure \ref{ratio_binned}, the normalization 
radii are $21\farcs85$, $3\farcs42$ and $1\farcs06$ for the JCMT, the BIMA and the SMA data, respectively. 
The force balance ($\frac{\sin\psi}{\sin\alpha}=\Sigma_B\equiv 1$) is indicated with the 
horizontal dashed black line. The effectively acting gravitational force (as 
introduced in Equation (\ref{eff_g})) is derived with the gravity efficiency
$<\epsilon_G>_r=1-\left<\frac{\sin\psi}{\sin\alpha}\right>_r$, which is 
symbolized through the lengths of the down-arrows. Similarly, the gravity 
dilution (by the magnetic field) is indicated with the dashed up-arrows.
The error bars, identical to the ones in Figure \ref{ratio_binned}, are omitted here.
}
\end{center}
\end{figure}

\begin{figure} 
\begin{center} 
\includegraphics[scale=0.7]{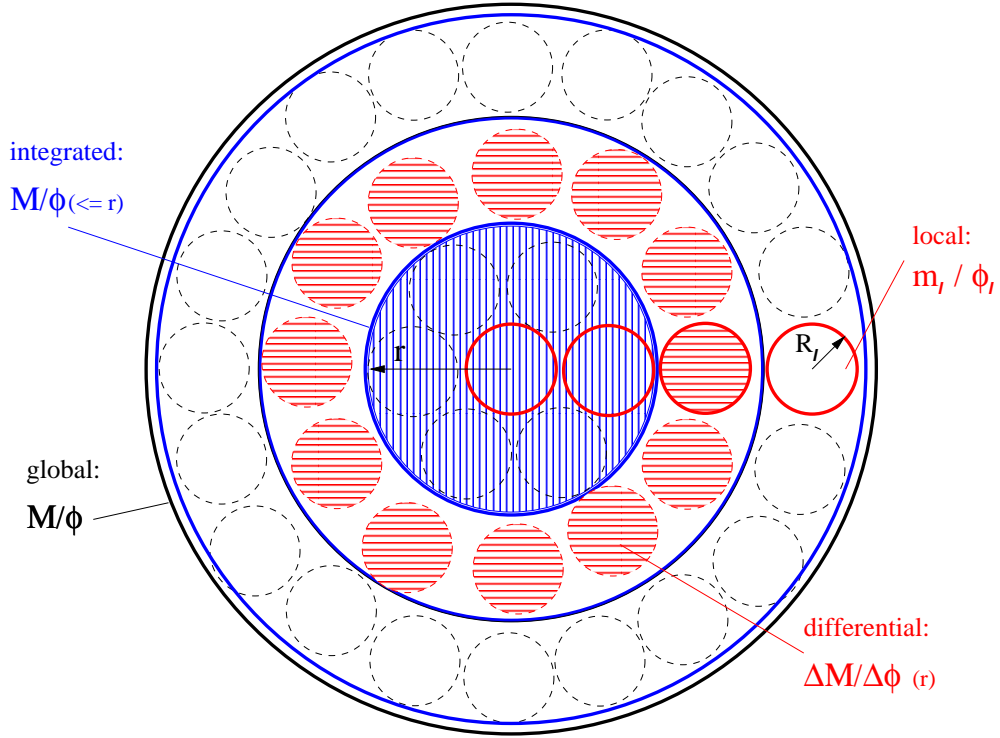}
 \caption{\label{m_f_ratio_schematic}
Illustration of the various mass-to-flux ratios as derived in 
Section \ref{m_f_ratio_discussion}.
Every small circle of radius $R_{\ell}$ represents a local mass-to-flux ratio
($\frac{m_{\ell}}{\phi_{\ell}}$, Equation (\ref{mf_mass})).
Its size is determined by the beam resolution.
The differential mass-to-flux ratio ($\frac{\Delta M}{\Delta \Phi}(r)$, Equation (\ref{diff_mf}))
at radius $r$ is calculated by azimuthally averaging all the local ratios of the 
hatched red circles inside the ring at radius $r$. The four solid red circles
symbolize differential ratios at four different radii.
The integrated mass-to-flux ratio ($\frac{M}{\phi}(\le r)$, Equation (\ref{mf_int}))
within radius $r$ is shown with the blue hatched circle. 
Concentric blue circles with growing radii symbolize integrated ratios
for increasingly larger volumes of the cloud. Azimuthally averaged ratios for 
each radius are displayed in the profiles in Figure \ref{m_f_ratio}.
A single global ratio ($\frac{M}{\phi}$, Equation (\ref{mf_glob})) is 
calculated from the black circle encompassing the entire cloud.
}
\end{center}
\end{figure}

\begin{figure} 
\begin{center} 
\includegraphics[scale=0.8]{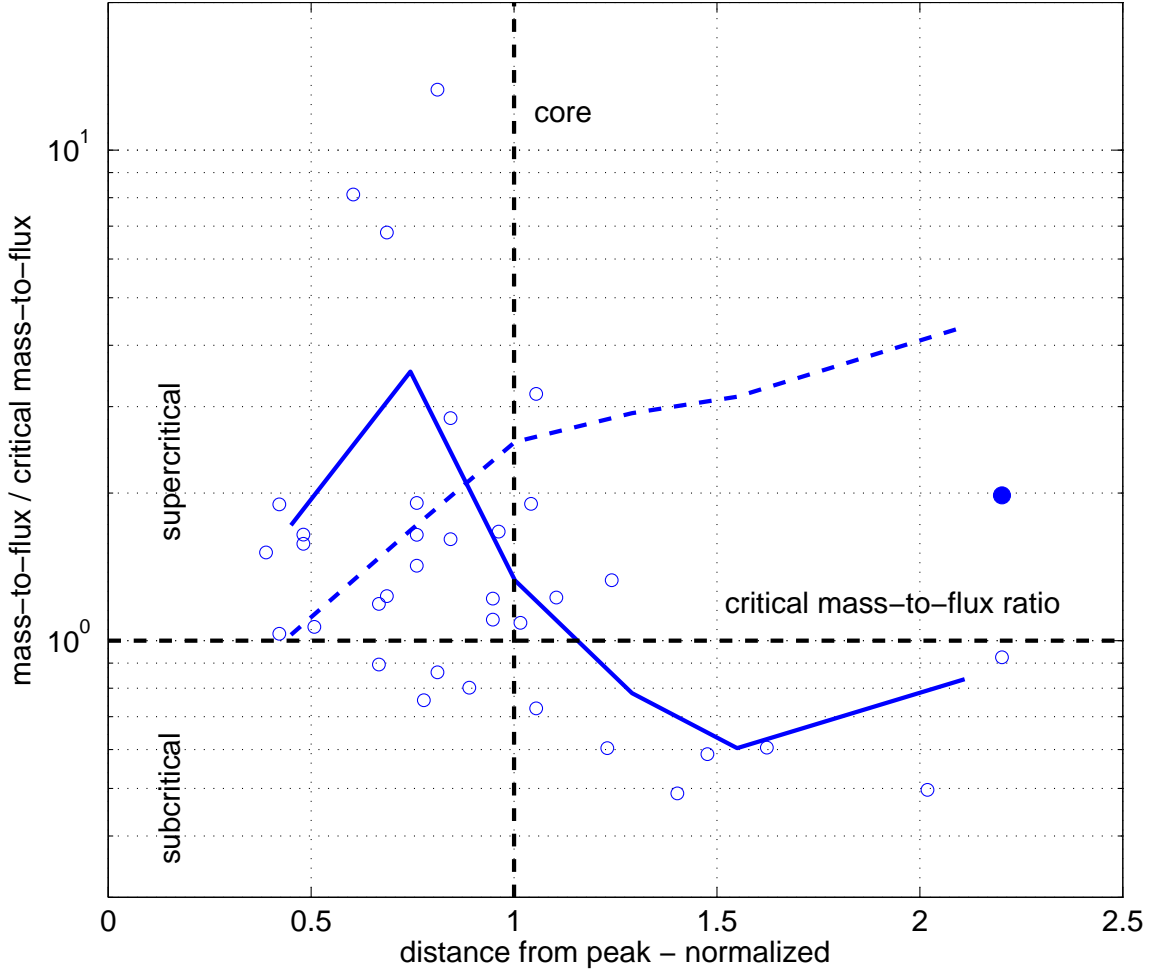}
 \caption{\label{m_f_ratio}\footnotesize
The various mass-to-flux ratios normalized to the corresponding critical 
mass-to-flux ratios (Section \ref{m_f_ratio_discussion}) for W51 e2. 
The ratios are shown
as a function of distance from the peak normalized to the core size.
The solid line displays the normalized azimuthally averaged differential 
mass-to-flux ratios (Equation (\ref{diff_mf}), converted from the magnetic field
to gravitational force ratios (Figure \ref{ratio_binned}) with the 
additional mass ratio, $\left(\frac{m_{\ell}(r)}{\bar{m_{\ell}}(r)}\right)^{1/2}\le 1$.
The mass ratio is derived from the integrated emission profiles in Figure \ref{ratio_binned}.
A clear transition between magnetically supercritical in the center to 
subcritical at larger distances is revealed. Open circles correspond to the individual ratios
before averaging and before multiplying with the mass ratio. Thus, they are local upper limits
(Equation (\ref{diff_mf})).
The dashed line is the integrated mass-to-flux ratio (Equation (\ref{mf_int})) normalized 
to the critical mass-to-flux ratio with its radius corresponding to each binning radius.
The single global mass-to-flux ratio (Equation (\ref{mf_glob})), averaging all force ratios
and normalizing to the core radius, is displayed with the filled circle.
Indicated with black dashed lines are the critical mass-to-flux ratio and the core size
(normalized to one).
When propagating uncertainties through 
$\frac{\Delta M}{\Delta \Phi}\sim \left(\frac{\sin\psi}{\sin\alpha}\right)^{-1/2}$,
errors are suspect to an additional factor
$\frac{1}{2}\left(\frac{\sin\psi}{\sin\alpha}\right)^{-3/2}$ as compared
to the force ratios (Figure \ref{ratio_binned}).
Thus, at larger radii where the force ratio is around one, errors in the 
mass-to-flux ratios are similar to or smaller than 
the ones for the field-to-gravity force, i.e. $\simlt \pm 10$\%. 
As the force ratios drop in the center area, errors grow larger by
a factor of about 2 to 3, i.e. reaching $\sim \pm 20$\% to $\sim \pm 30$\%.
For clarity, error bars are not displayed here.
}
\end{center}
\end{figure}

\begin{figure} 
\begin{center} 
\includegraphics[scale=0.8]{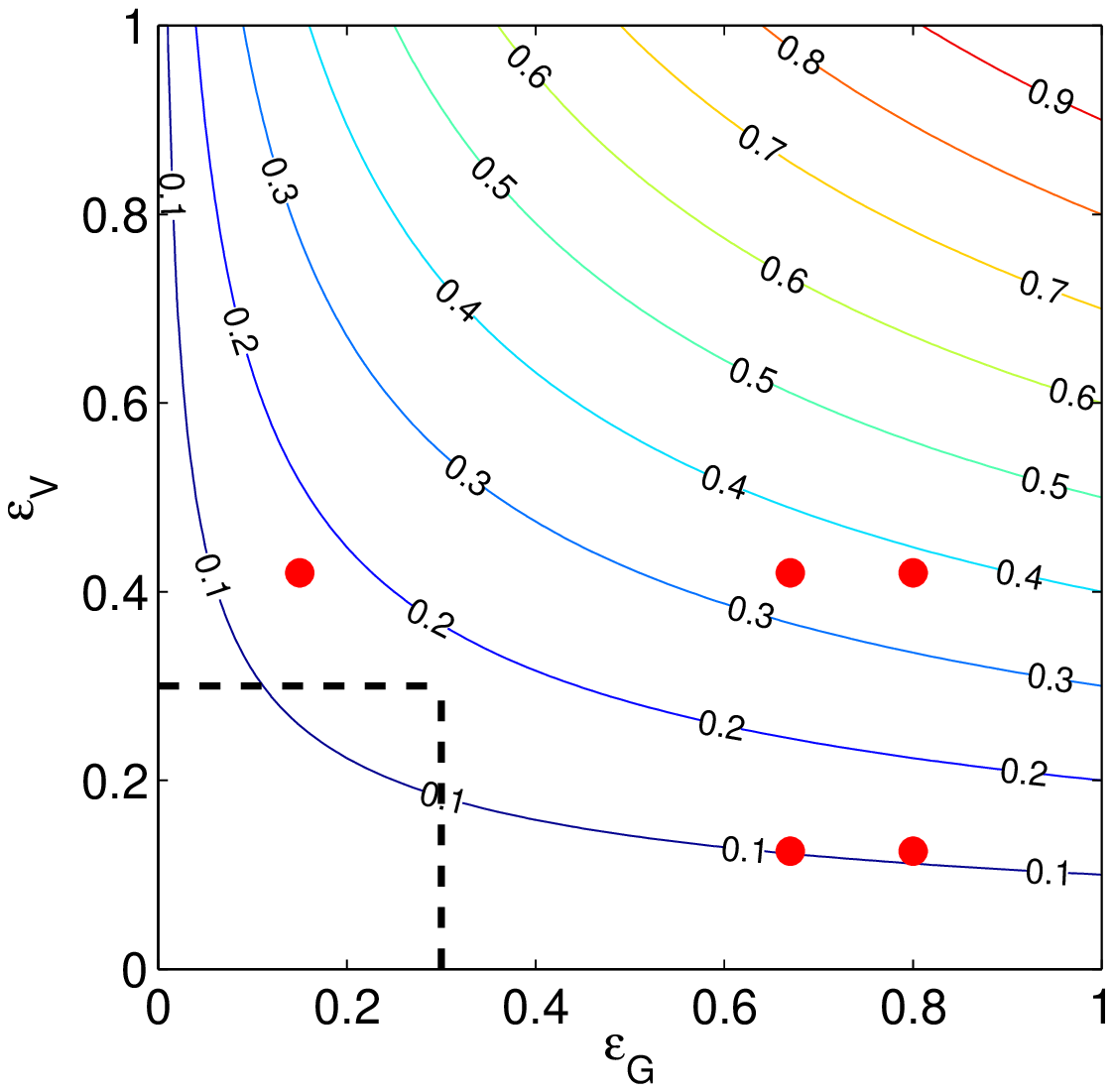}
\includegraphics[scale=0.8]{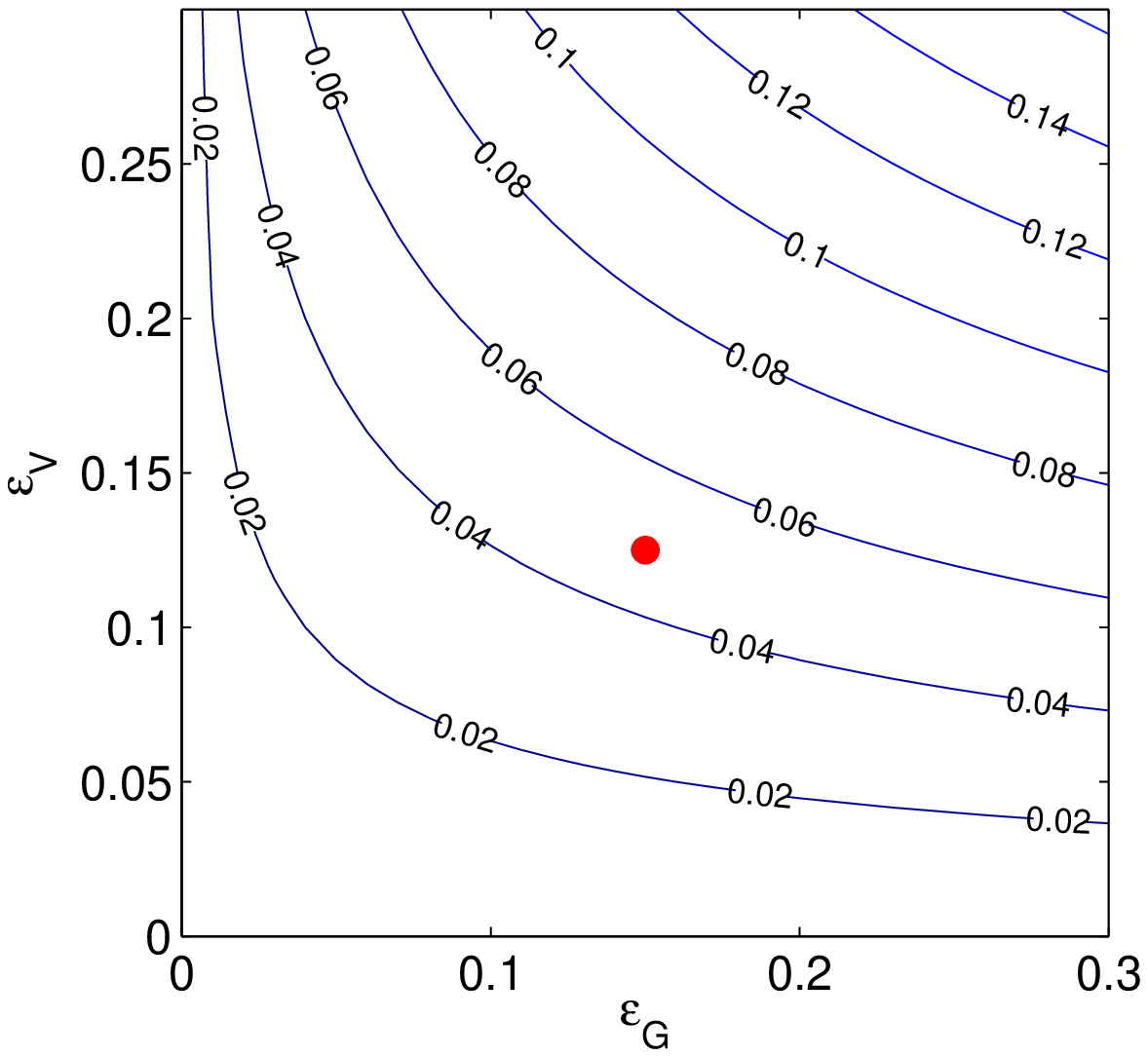}
 \caption{\label{sf_efficiency}
Estimated reduced star formation efficiency (contour levels)
with a diluted gravitational
force (efficiency factor $\epsilon_G$) and a reduced collapsing volume (efficiency factor 
$\epsilon_V$) following Equation (\ref{eq_sf_eff}). 
The lower panel shows the zoomed-in area marked with dashed lines in the upper panel.
The red dots are possible estimated efficiencies based on the force ratio 
(Figure \ref{ratio_universal}) and the 
differential mass-to-flux ratio (Figure \ref{m_f_ratio}) for the case of W51 e2.
}
\end{center}
\end{figure}

\begin{enumerate}

\item
{\it Magnetic field - intensity gradient correlation:} 
The correlation in the magnetic field and dust intensity gradient orientations --
which served as a starting point for the new method developed in \citet{koch11} 
-- is also found at larger scales in the BIMA and JCMT data.
The correlation coefficients seem to be the larger the higher the resolution
is; i.e. the tightest correlation is found for the collapsing core e2
(Figure \ref{map_b_i}, left panels).

\item
{\it Magnetic Field Significance:}
Maps of the magnetic field to gravity force ratio show distinct features for
all three data sets (Figure \ref{map_b_i}, right panels). At larger distances
from the emission peaks, the field force is comparable or larger than the 
gravity force (ratio $\simgt 1$). Closer to the emission peaks, gravity 
becomes more dominant (ratio $<1$). This suggests and quantifies that gravity 
is not everywhere acting equally efficiently, but is being more significantly 
opposed by the magnetic field tension force at larger distances. 
Based on this finding, we introduce a gravity
efficiency factor. This also seems to be in agreement with the concept of
gravity dilution (Figure \ref{ratio_universal}). 
Moreover, this is in support of inside-out collapse scenarios.

\item
{\it Self-similarity:}
Azimuthally averaged radial profiles of the force ratios show similar 
features for the JCMT, BIMA and the SMA data (Figure \ref{ratio_binned}).
When normalized to core sizes, these profiles closely align (Figure \ref{ratio_universal}).
This points toward self-similar properties in the magnetic field and gravity
interplay from the large parent cloud down to the collapsing core.

\item
{\it Mass-to-flux ratio:}
The force ratio can be converted into a mass-to-flux ratio. In its most general 
form, the mass-to-flux ratio can then be expressed as the inverse square root
of the force ratio multiplied by the square root of a ratio of two mass elements.
With the force ratio being a local criterion, this naturally leads to a local
(differential) mass-to-flux ratio for a subvolume in a molecular cloud as a
function of position. An integrated and global mass-to-flux ratio can be derived
by appropriately adding and averaging local quantities. Similar to the force
ratio, all the various mass-to-flux ratios do not have to rely on absolute 
mass and field strength inputs, but they are largely model-independent. This 
finding then does not only provide a criterion to decide whether a molecular 
cloud as an entity is magnetically supercritical or not, but it extends
this criterion to any subvolume (Figures \ref{m_f_ratio_schematic} and \ref{m_f_ratio}).
In the case of W51 e2, a transition occurs from subcritical at larger distances
to supercritical in the central area. This might explain why fragmentation
is found to happen preferentially in the center.

\item
{\it Star formation efficiency:}
A reduced star formation efficiency (compared to the non-magnetized free-fall
case) is derived based on two observed magnetic field properties: (1) Gravity 
dilution (Figure \ref{ratio_universal}) defines an effectively available gravity
force to initiate and drive a collapse and, therefore, it increases the collapse
time. (2) The differential mass-to-flux ratio (as a function of radius) provides
a hint that the magnetic field reduces the volume of supercritical state and, 
therefore, only a limited volume can collapse and form stars (Figure \ref{m_f_ratio}).
With these two findings a modified accretion rate can be calculated. Comparing this
to a standard free-fall rate shows that the magnetic field is capable of reducing
a standard star formation efficiency to 10\% or less.

\end{enumerate}

The authors thank the referee for valuable comments which led to further
important insight in this work.
P.T.P.H. is supported by NSC grant NSC97-2112-M-001-007-MY3.




\begin{thebibliography}{}

\bibitem[Allen et al.(2003)]{allen03} Allen, A., Li, Z.-Y., \& Shu, F. H. 2003, \apj, 599, 363

\bibitem[Bertram et al.(2011)]{bertram11} Bertram, E., Federrath, C., Banerjee, R., \& Klessen, R.S. 2011, 
                             arXiv:1111.5539v1

\bibitem[Chrysostomou et al.(2002)]{chrys02} Chrysostomou, A., Aitken,
D.K., Jenness, T., et al. 
2002, A\&A, 385, 1014

\bibitem[Crutcher et al.(2009)]{crutcher09} Crutcher, R.M., Hakobian, N.,
\& Troland, T.H. 2009, \apj, 692, 844

\bibitem[Dotson et al.(2010)]{dotson10} 
Dotson, J.L., Vaillancourt, J.E., Kirby, L., et al. 2010, ApJS, 186, 406

\bibitem[Draine \& Weingartner(1996)]{draine96} Draine, B.T., \&
Weingartner, J.C. 1996, \apj, 470, 551

\bibitem[Draine \& Weingartner(1997)]{draine97} Draine, B.T., \& 
Weingartner, J.C. 1997, \apj, 480, 633

\bibitem[Falceta-Gon\c{c}alves et al.(2008)]{falceta08}
Falceta-Gon\c{c}alves, D., Lazarian, A., \&
                                           Kowal, G. 2008, \apj, 679, 537

\bibitem[Genzel et al.(1981)]{genzel81} Genzel, R., Downes, D., Schneps, M.H., et al. 1981, \apj,
247, 1039

\bibitem[Girart et al.(2006)]{girart06} Girart, J.M., Rao, R., \&
Marrone, D.P. 2006, Science, 313, 812

\bibitem[Girart et al.(2009)]{girart09} Girart, J.M., Beltr\'an, M. T., Zhang, Q., 
Rao, R., \&  Estalella, R. 2009, Science, 324, 1408

\bibitem[Hezareh et al.(2011)]{hezareh11} Hezareh, T., Houde, M., McCoey, C., \& Li, H. 2011, arXiv:1007.2242

\bibitem[Hildebrand(1988)]{hildebrand88} Hildebrand, R.H. 1988, QJRAS, 29, 327

\bibitem[Hildebrand et al.(2009)]{hildebrand09} Hildebrand, R.H., Kirby,
L., Dotson, J.L., Houde, M., Vaillancourt, J.E. 2009,
\apj, 696, 567

\bibitem[Ho \& Young(1996)]{ho96} Ho, P.T.P., \& Young, L.M. 1996, \apj, 472, 742

\bibitem[Ho et al.(2004)]{ho04} Ho, P.T.P., Moran, J.M., \& Lo,
K.-Y. 2004, \apj, 616, 1

\bibitem[Houde et al.(2009)]{houde09} 
Houde, M., Vaillancourt, J.E., Hildebrand, R.H., 
                               Chitsazzadeh, S., \& Kirby, L. 2009, \apj, 706, 1504

\bibitem[Koch et al.(2010)]{koch10}Koch, P.M., Tang, Y.-W., \& Ho, P.T.P. 2010, \apj, 
721, 815

\bibitem[Koch et al.(2012)]{koch11}Koch, P.M., Tang, Y.-W., \& Ho, P.T.P. 2012, \apj, accepted

\bibitem[Krumholz \& Tan(2007)]{krumholz07} Krumholz, M. R. \& Tan, J. C., 2007, \apj, 656, 959

\bibitem[Lai et al.(2001)]{lai01} Lai, S.-P., Crutcher, R.M., Girart,
J.M., \& Rao, R. 2001, \apj, 561, 864L

\bibitem[Li \& Houde(2008)]{li08} Li, H., \& Houde, M. 2008, \apj, 677, 1151

\bibitem[Li \& Nakamura(2004)]{li04} Li, Z.-Y., \& Nakamura, F. 2004, \apj, 609, L83

\bibitem[Mestel \& Spitzer(1956)]{mestel56} Mestel, L., \& Spitzer, L. 1956, MNRAS, 116, 505 

\bibitem[Nakamura \& Li(2011)]{nakamura11} Nakamura, F., \&  Li, Z.-Y. 2011, arXiv:1107.3616

\bibitem[Lazarian(2000)]{lazarian00} Lazarian, A. 2000, ASPC, 215, 69

\bibitem[Mouschovias(1976)]{mouschovias76} Mouschovias, T.Ch. 1976, \apj, 207, 141

\bibitem[Rao et al.(2009)]{rao09} 
Rao, R., Girart, J.M., Marrone, D.P., Lai, S.-P., \& Schnee, S. 2009, \apj, 707, 921

\bibitem[Rudolph et al.(1990)]{rudolph90} Rudolph, A., Welch, W. J., Palmer, P., \& Dubrulle, B. 1990, 
                                                      \apj, 363, 528

\bibitem[Solins et al.(2004)]{solins04} Sollins, P. K., Zhang, Q., \& Ho, P.T.P., 2004, \apj, 606, 943

\bibitem[Surcis et al.(2011)]{surcis11} Surcis, G.,  Vlemmings, W.H.T., Curiel, S., et al.
                                    2011, A\&A, 527A, 48S

\bibitem[Shi et al.(2010)]{shi10} Shi, H., Zhao, J.-H., \& Han, J.L. 2010, \apj, 718, L181

\bibitem[Shu \& Li(1997)]{shu97}  Shu, F.H., \& Li, Z.-Y. 1997, \apj, 475, 251

\bibitem[Shu et al.(1999)]{shu99} Shu, F.H., Allen, A., Shang, H., Ostriker, E.C. 
                                   \& Li, Z.-Y. 1999, in 
The Origin of Stars and Planetary Systems, edited by Charles J. Lada and Nikolaos D. Kylafis, 
Kluwer Academic Publishers, 1999, p.193

\bibitem[Shu et al.(2004)]{shu04} Shu, F.H., Li, Z.-Y., \& Allen, A. 2004, \apj, 601, 930

\bibitem[Tang et al.(2009a)]{tang09a} Tang, Y.-W., Ho, P.T.P., Girart,
J.M., et al. 
                                      2009a, \apj, 695, 1399

\bibitem[Tang et al.(2009b)]{tang09b} Tang, Y.-W., Ho, P.T.P., Koch, P.M., et al.
                                      2009b, \apj, 700, 251

\bibitem[Tang et al.(2010)]{tang10} Tang, Y.-W., Ho, P.T.P., Koch, P.M., \&
                                      Rao, R. 2010, \apj, 717, 1262

\bibitem[V\'azquez-Semadeni et al.(2011)]{vazquez11} V\'azquez-Semadeni, E., Banerjee, R., G\'omez, G.C., 
                                                    et al. 2011, MNRAS, 414, 2511

\bibitem[Vlemmings et al.(2011)]{vlemmings11} Vlemmings, W.H.T., Humphreys, E.M.L., \& Franco-Hern\'andez, R. 2011, \apj, 728, 149

\bibitem[Young et al.(1998)]{young98} Young, L. M., Keto, E., \& Ho, P.T.P., 1998, \apj, 507, 270

\bibitem[Zhang \& Ho(1997)]{zhang97} Zhang, Q., \& Ho, P.T.P., 1997, \apj, 488, 241

\bibitem[Zhang et al.(1998)]{zhang98} Zhang, Q., Ho, P.T.P., \& Ohashi, H., 1998, \apj, 494, 636


\end{thebibliography}
\end{document}